%% file: TOP-16-003_temp.tex
\pdfoutput=1

\documentclass[11pt,twoside,a4paper,cmspaper,final,collab]{cms-tdr}
\def\svnVersion{426958P}\def\cmsCernNoTag{CERN-EP-2016-233}\def\cmsCernDate{\today}\def\cmsMessage{Published in Physics Letters B as \href{http://dx.doi.org/10.1016/j.physletb.2017.07.047}{\doi{10.1016/j.physletb.2017.07.047.}}}

\begin{document}\cmsNoteHeader{TOP-16-003}

\hyphenation{had-ron-i-za-tion}
\hyphenation{cal-or-i-me-ter}
\hyphenation{de-vices}
\RCS$Revision: 426908 $
\RCS$HeadURL: svn+ssh://svn.cern.ch/reps/tdr2/papers/TOP-16-003/trunk/TOP-16-003.tex $
\RCS$Id: TOP-16-003.tex 426908 2017-09-26 16:06:31Z alverson $
\newlength\cmsFigWidth
\ifthenelse{\boolean{cms@external}}{\setlength\cmsFigWidth{0.95\columnwidth}}{\setlength\cmsFigWidth{0.7\textwidth}}
\ifthenelse{\boolean{cms@external}}{\providecommand{\cmsLeft}{top\xspace}}{\providecommand{\cmsLeft}{left\xspace}}
\ifthenelse{\boolean{cms@external}}{\providecommand{\cmsRight}{bottom\xspace}}{\providecommand{\cmsRight}{right\xspace}}

\newcommand{\vtb}{\ensuremath{V_{\mathrm{tb}}}\xspace}
\newcommand{\PFrelIso}{\ensuremath{I_{\text{rel}}}\xspace}
\newcommand{\pfPhotonIso}{\ensuremath{I^{\gamma}}\xspace}
\newcommand{\pfChargedHadronIso}{\ensuremath{I^{\text{ch. h}}}\xspace}
\newcommand{\pfNeutralHadronIso}{\ensuremath{I^{\text{n. h}}}\xspace}
\newcommand{\pfPU}{\ensuremath{I^{\text{PU ch. h}}\xspace}}
\newcommand{\mTW}{\ensuremath{m_{\mathrm{T}}^{\PW}}\xspace}
\newcommand{\mW}{\ensuremath{m_{\PW}}\xspace}
\newcommand{\wjets}{\ensuremath{\PW+\text{jets}}\xspace}
\newcommand{\zjets}{\ensuremath{\Z+\text{jets}}\xspace}
\newcommand{\mptvec}{\ensuremath{{\vec{p}}_\mathrm{T}\hspace{-1.02em}/\kern 0.5em}\xspace}
\newcommand{\mpx}{\ensuremath{{p}_{x}\hspace{-1.02em}/\kern 0.5em}\xspace}
\newcommand{\mpy}{\ensuremath{{p}_{y}\hspace{-1.02em}/\kern 0.5em}\xspace}
\newcommand{\sigmat}{\ensuremath{\sigma_{t\text{-ch.},\PQt+\PAQt}}\xspace}
\newcommand{\sigmattop}{\ensuremath{\sigma_{t\text{-ch.,\PQt}}}\xspace}
\newcommand{\sigmatantitop}{\ensuremath{\sigma_{t\text{-ch.,}\PAQt}}\xspace}
\newcommand{\xsectheotop}{ 136.02 \xspace}
\newcommand{\xsectheoantitop}{ 80.95 \xspace}
\newcommand{\xsectheo}{ 216.99 \xspace}
\renewcommand{\xsectheotop}{ 136.0 \xspace}
\renewcommand{\xsectheoantitop}{ 81.0 \xspace}
\renewcommand{\xsectheo}{ 217.0 \xspace}
\newcommand{\xsectheotopscale}{\ensuremath{^{+4.09}_{-2.92}}\xspace}
\newcommand{\xsectheoantitopscale}{\ensuremath{^{+2.53}_{-1.71}}\xspace}
\newcommand{\xsectheoscale}{\ensuremath{^{+6.62}_{-4.64}}\xspace}
\renewcommand{\xsectheotopscale}{\ensuremath{^{+4.1}_{-2.9}}\xspace}
\renewcommand{\xsectheoantitopscale}{\ensuremath{^{+2.5}_{-1.7}}\xspace}
\renewcommand{\xsectheoscale}{\ensuremath{^{+6.6}_{-4.6}}\xspace}
\newcommand{\xsectheotoppdf}{\ensuremath{\pm3.52}\xspace}
\newcommand{\xsectheoantitoppdf}{\ensuremath{\pm3.18}\xspace}
\newcommand{\xsectheopdf}{\ensuremath{\pm6.16}\xspace}
\renewcommand{\xsectheotoppdf}{\ensuremath{\pm3.5}\xspace}
\renewcommand{\xsectheoantitoppdf}{\ensuremath{\pm3.2}\xspace}
\renewcommand{\xsectheopdf}{\ensuremath{\pm6.2}\xspace}
\newcommand{\vtbobs}{ 1.05 \xspace}
\newcommand{\vtbobsexp}{ 0.07 \xspace}
\newcommand{\vtbobstheo}{ 0.02 \xspace}
\newcommand{\lumiunc}{2.3\%\xspace}
\newcommand{\twoJoneT}{2-jets--1-tag\xspace}
\newcommand{\twoJzeroT}{2-jets--0-tag\xspace}
\newcommand{\threeJoneT}{3-jets--1-tag\xspace}
\newcommand{\threeJtwoT}{3-jets--2-tags\xspace}
\newcommand{\nJmT}{``$n$-jets--$m$-tag(s)"\xspace}
\newcommand{\x}{\ensuremath{\phantom{0}}}

\cmsNoteHeader{TOP-16-003}

\title{Cross section measurement of $t$-channel single top quark production in
pp collisions at $\sqrt{s} = 13\TeV$}

\date{\today}
\abstract{The cross section for the production of single top quarks in the $t$
channel is measured in proton-proton collisions at 13\TeV with the CMS detector at the LHC. The analyzed data correspond to an integrated luminosity of 2.2\fbinv. The event selection requires one muon and two jets where one of the jets is identified as originating from a bottom quark. Several kinematic variables are then combined into a multivariate discriminator to distinguish signal from background events.  A fit to the distribution of the discriminating variable yields a total cross section of $238 \pm 13\stat \pm 29\syst\unit{pb}$ and a ratio of top
quark and top antiquark production of $R_{t\text{-ch.}}=1.81 \pm
0.18\stat \pm 0.15\syst $. From the total cross section
the absolute value of the CKM matrix element $V_{\PQt\PQb}$ is calculated to be $1.05 \pm0.07\,\text{(exp)}\pm0.02\thy$. All results are in agreement with the standard model predictions.}

\hypersetup{%
pdfauthor={CMS Collaboration},%
pdftitle= {Cross section measurement of t-channel single top quark production
in  pp collisions at sqrt(s) = 13 TeV},%
pdfsubject={CMS},%
pdfkeywords={CMS, physics, top quark}}
\maketitle

\section{Introduction}
\label{sec:introduction}
\begin{figure}[h]
\begin{center}
\includegraphics[width=\textwidth]{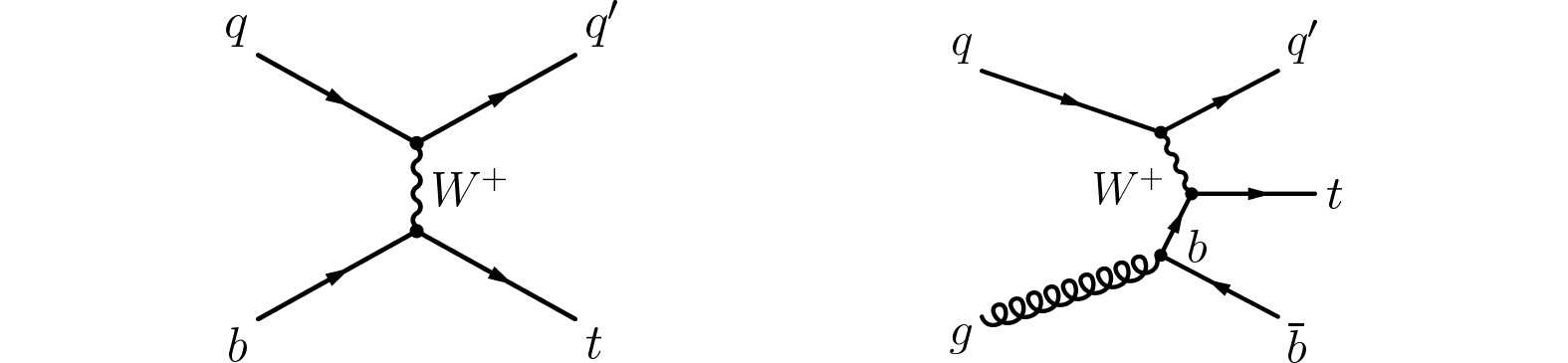}
\caption{\label{fig:FG} Feynman diagrams for single top quark production in the $t$ channel: (\cmsLeft) 2$\to$2 and (\cmsRight) 2$\to$3 processes.}
\end{center}

\end{figure}

The production of single top quarks provides a unique testing ground for the study of electroweak processes, specifically the tWb vertex, as well as the measurement of the Cabibbo--Kobayashi--Maskawa (CKM) matrix element \vtb. The single top quark production was first detected at the Tevatron~\cite{Abazov:2009ii,Aaltonen:2009jj} and was studied at higher energies~\cite{Chatrchyan:2012ep,Aad:2012ux,PASSingleTopCrossSection,PhysRevLett.112.231802} at the CERN LHC~\cite{1748-0221-3-08-S08001}. At the LHC, the dominant production mechanism of single top quarks is the $t$-channel process. The other two processes, W-associated (tW) production and production via the $s$ channel, amount to roughly 30\% of the total single top quark production cross section at 13\TeV~\cite{Kidonakis:2012db}. The $t$-channel production mode, presented in Fig.~\ref{fig:FG}, has a very distinct signature because of the presence, within the detector acceptance, of a light quark recoiling against the top quark. The CMS collaboration has performed several measurements of this process using data collected at $\sqrt{s} = 7$ and 8\TeV~\cite{CMS-PAS-TOP-13-001,Khachatryan:2014vma,PASSingleTopCrossSection}.
This analysis is based on a data set obtained from proton-proton collisions at a centre-of-mass energy of 13\TeV, corresponding to an integrated luminosity of 2.2\fbinv.
The cross section calculation of $t$-channel single top quark production can be performed in two different schemes~\cite{fourfiveflavorschemescamp, fourfiveflavorschemesmalt, fourfiveflavorschemes}. In the five-flavour scheme (5FS) b quarks come from the incoming proton and the leading order (LO) diagram is a 2$\to$2 process (Fig.~\ref{fig:FG}\,\cmsLeft), while in the four-flavour scheme (4FS) b quarks are not present in the initial state, and the LO diagrams are 2$\to$3 processes (Fig.~\ref{fig:FG}\,\cmsRight).

The next-to-leading-order (NLO) calculations with \textsc{Hathor}\,v2.1~\cite{HATOR,Kant:2014oha} in the 5FS result in cross section values of
\begin{linenomath}
\begin{align*}
\sigmattop &=  \xsectheotop \, \xsectheotopscale\,\text{(scale)} \xsectheotoppdf \,(\mathrm{PDF}{+}{\alpha_\mathrm{S}})\unit{pb},\\
\sigmatantitop &=   \x \xsectheoantitop \, \xsectheoantitopscale\,\text{(scale)}\xsectheoantitoppdf\,(\mathrm{PDF}{+}{\alpha_\mathrm{S}})\unit{pb},\\
\sigmat &=  \xsectheo \, \xsectheoscale\, \text{(scale)} \xsectheopdf \,(\mathrm{PDF}{+}{\alpha_\mathrm{S}})\unit{pb},
\end{align*}
\end{linenomath}
for the $t$-channel production at $\sqrt{s}=13\TeV$ of a top quark, antiquark, and the sum, respectively.
The above cross sections are evaluated for a top quark mass of 172.5\GeV, using the PDF4LHC prescription~\cite{Botje:2011sn} for  the parton distribution functions (PDFs). The uncertainties are associated with the renormalization and factorization scale uncertainty as well as the PDF and $\alpha_\mathrm{S}$ uncertainties which are calculated with the MSTW2008 68\% CL NLO~\cite{MSTW2008NLO,Martin:2009bu}, CT10 NLO~\cite{CT10}, and NNPDF2.3~\cite{Ball:2012cx} PDF sets.
Calculations at next-to-next-to-leading order (NNLO)~\cite{NNLO-Caola} are expected to be different from NLO by only a few percent. Similar results are obtained at NLO as a function of the centre-of-mass energy with next-to-next-to-leading logarithms (NNLL) considered~\cite{Kidonakis:2011wy}.
In the analysis described in this letter, the separation between signal and background processes is achieved using a multivariate analysis (MVA) technique. An artificial neural network is employed to construct a single classifier, exploiting the discriminating power of several kinematic distributions.
The cross section of $t$-channel single top quark production is determined from a fit to the distribution of this single variable. Events with an isolated muon in the final state are selected; the muon originates from the decay of the W boson from the top quark, either directly or through $\PW\to\tau\nu$ decays. No attempts are made to distinguish these two cases and the signal yield is corrected for the $\tau$ decay contributions using the corresponding theoretical branching ratio.

\section{The CMS detector and the simulation of events}
\label{sec:CMSdetector}
The central feature of the CMS apparatus is a superconducting solenoid of 6\unit{m} internal diameter, providing a magnetic field of 3.8\unit{T}. Within the solenoid volume are a silicon pixel and strip tracker, a lead tungstate crystal electromagnetic calorimeter (ECAL), and a brass and scintillator hadron calorimeter (HCAL), each composed of a barrel and two endcap sections. Forward calorimeters extend the pseudorapidity ($\eta$) ~\cite{Chatrchyan:2008zzk} coverage provided by the barrel and endcap detectors. Muons are measured in the range $\abs{\eta} < 2.4$ using gas-ionization detectors embedded in the steel flux-return yoke outside the solenoid. Matching muons to tracks measured in the silicon tracker results in a relative transverse momentum ($\pt$) resolution for muons with $20 <\pt < 100\GeV$ of 1.3--2.0\% in the barrel and better than 6\% in the endcaps. The \pt resolution in the barrel is better than 10\% for muons with \pt up to 1\TeV~\cite{Chatrchyan:2012xi}. A more detailed description of the CMS detector, together with a definition of the coordinate system used and the relevant kinematic variables, can be found in Ref.~\cite{Chatrchyan:2008zzk}.
Monte Carlo (MC) simulation event generators are used to create simulated signal and background samples. Signal $t$-channel events are generated at NLO with {\MADGRAPH}\_a\MCATNLO version 2.2.2 (MG5\_a\MCATNLO)~\cite{amcatnlo} in the 4FS. The $\ttbar$ and tW background processes are generated with \POWHEG~2.0~\cite{Re:2010bp,Alioli:2010xd,Alioli:2009je,Frixione:2007vw}. The latter is simulated in the 5FS. The value of the top quark mass used in the simulated samples is $m_{\cPqt}=172.5\GeV$. For all samples {\PYTHIA~8.180}~\cite{Sjostrand:2006za} with tune CUETP8M1~\cite{Skands:2014pea} is used to simulate the parton shower, hadronization, and the underlying event. Simulated event samples with W and Z bosons in
association with jets are generated using MG5\_a\MCATNLO and the FxFx merging scheme~\cite{Frederix:2012ps}, where up to two additional partons are generated at the matrix-element level. The quantum chromodynamics (QCD) multijet events, generated with {\PYTHIA~8.180}, are used to validate the estimation of this background with a technique based on control samples in data. The default parametrization of the PDF used in all simulations is NNPDF30\_nlo\_as\_0118~\cite{NNPDF30}. All generated events undergo a full simulation of the detector response according to the implementation of the CMS detector within {\GEANTfour}~\cite{geant}. Additional proton-proton interactions within the same or nearby bunch crossing (pileup) are included in the simulation with the same distribution as observed in data.

\section{Event selection and reconstruction}
\label{sec:selection}
Events with exactly one muon and at least two jets are considered in this analysis. In addition to the presence of exactly one isolated muon, the signature of $t$-channel single top quark production is characterized by a substantial momentum imbalance associated to at least one neutrino, a jet arising from the hadronization of a bottom quark (b jet) from the top quark decay, and a light-quark jet --- often produced in the forward region. Some events also feature a second b jet, coming from the second b quark in the gluon splitting (as shown in Fig.~\ref{fig:FG}\,\cmsRight). This second b jet is often not selected for the analysis as the $\pt$ spectrum is generally softer and broader than that of the b jet from the top quark decay. To select events for further analysis, a high-level trigger (HLT) that requires the presence of an isolated muon with $\pt > 20\GeV$ is used. From the sample of triggered events, only those with at least one primary vertex reconstructed from at least four tracks, with the longitudinal (radial) distance of less than 24 (2)$\,$cm from the centre of the detector, are considered for the analysis. Among all primary vertices in the event, the one with the largest scalar sum of $\pt^{2}$ of associated particles is selected.
The particle flow (PF) algorithm~\cite{CMS-PAS-PFT-09-001, CMS-PAS-PFT-10-001} is used to reconstruct and identify individual particles in the event using combined information from the various subdetectors of the CMS experiment.
Muon candidates are reconstructed combining the information from both the silicon tracker and the muon spectrometer in a global fit. An identification is performed using the quality of the geometrical matching between the tracker and the muon system measurements. The transverse momentum of muons is obtained from the curvature of the corresponding tracks. The energy of photons is directly obtained from the ECAL measurement, corrected for zero-suppression effects. The energy of electrons is determined from a combination of the electron momentum at the primary interaction vertex determined by the tracker, the energy of the corresponding ECAL cluster, and the energy sum of all bremsstrahlung photons spatially compatible with originating from the electron track. The energy of charged hadrons is determined from a combination of their momenta measured in the tracker and the matching ECAL and HCAL energy deposits, corrected for zero-suppression effects and for the response function of the calorimeters to hadronic showers. Finally, the energy of neutral hadrons is obtained from the corresponding corrected ECAL and HCAL energy.
Using this information, the muon isolation variable, \PFrelIso, is defined as
\begin{linenomath}
\begin{equation}
\PFrelIso = \frac{ \pfChargedHadronIso + \max [(\pfPhotonIso + \pfNeutralHadronIso -  0.5 \times\pfPU),0]}{\pt},
\label{eq:pfreliso}
\end{equation}
\end{linenomath}
where $\pfChargedHadronIso$, $\pfPhotonIso$, $\pfNeutralHadronIso$, and $\pfPU$ are, respectively, the scalar $\pt$ sums of the charged hadrons, photons, neutral hadrons, and charged hadrons associated with pileup vertices. The sums are computed in a cone of $\Delta R\equiv\sqrt{\smash[b]{(\Delta\eta)^{2}+(\Delta\phi)^{2}}} = 0.4$ around the muon direction, where $\phi$ is the azimuthal angle in radians. The contribution $0.5 \times \pfPU$ accounts for the expected pileup contribution from neutral particles. It is determined from the measured scalar $\pt$ sum of charged hadrons $\pfPU$, corrected for the neutral-to-charged particle ratio as expected from isospin invariance.
Events are selected if they contain exactly one muon candidate with $\pt > 22\GeV$, $|\eta|< 2.1$, and $\PFrelIso < 0.06$. Events with additional muon or electron candidates, passing looser selection criteria, are rejected. The loose selection criteria are $\pt > 20~(10)\GeV$, $\abs{\eta}< 2.5$, and $\PFrelIso < 0.2$ for additional electrons (muons) where the electron isolation has a similar definition to that of the muon.
Jets are reconstructed by clustering PF particle candidates using the anti-$\kt$ clustering algorithm~\cite{Cacciari:2008gp} with a distance parameter of 0.4. Charged-particle candidates closer along the $z$ axis to any vertex other than the selected primary vertex are not included.
A correction to account for pileup interactions is estimated on an event-by-event basis using the jet area method described in Ref.~\cite{Cacciari:2007fd}, and is applied to the reconstructed jet \pt. Further jet energy corrections, derived from the study of dijet events and photon plus jet events in data, are applied. Jets are required to have $|\eta|<4.7$ and $\pt > 40\GeV$.
Once the jets have been selected according to the above criteria, they can be further categorized using a b tagging discriminator variable in order to distinguish between jets stemming from the hadronization of b quarks and those from the hadronization of light partons. The combined secondary vertex algorithm uses track-based lifetime information together with secondary vertices inside the jet to provide a MVA discriminator for b jet identification~\cite{1748-0221-8-04-P04013,CMS-PAS-BTV-15-001}. At the chosen working point, the efficiency of the tagging algorithm to correctly find b jets is about 45\% with a rate of 0.1\% for mistagging light-parton jets~\cite{1748-0221-8-04-P04013}.
Events are divided into categories according to the number of selected jets and b-tagged jets. In the following, categories are labelled as \nJmT, referring to events with $n$ jets, $m$ of which are identified as b jets. The category enriched in $t$-channel signal events is the \twoJoneT category, while the \threeJoneT and \threeJtwoT categories are enriched in \ttbar~background events and are used to constrain the \ttbar~contribution in the final fit. The \twoJzeroT category provides good sensitivity for the validation of the W+jets simulation.
To reject events from QCD multijet background processes, a requirement on the transverse mass of the W boson of $\mTW > 50\GeV$ is imposed, where
\begin{linenomath}
\begin{equation}
\mTW =\sqrt{\left(p_{\mathrm{T},\mu} + \PTslash\right)^2 - \left( p_{x,\mu} + \mpx \right)^2 - \left( p_{y,\mu} + \mpy \right)^2}.
\end{equation}
\end{linenomath}
Here, $\PTslash$ is defined as the magnitude of $\mptvec$ which is the negative of the vectorial $\pt$ sum of all the PF particles. The $\mpx$ and $\mpy$  quantities are the $\mptvec$ components along the $x$ and $y$ axes, respectively.
In Table~\ref{tab:yield}, the number of selected events is shown for the \twoJoneT signal region, separately for events with muons of positive and negative charge.
Except for the QCD multijet process, which is determined from a fit to data and presented with the corresponding systematic uncertainties, all simulated samples are normalized to the expected cross sections
with uncertainties corresponding to the size of the samples. The main backgrounds arise from $\bbbar$, W+jets, and QCD multijet processes.

\begin{table} [!h!tbp]
\topcaption{Event yields for the main processes in the \twoJoneT
sample. The quoted uncertainties are statistical only. All yields are taken from simulation, except for QCD multijet events where the yield and the associated uncertainty are determined from data (as discussed in Section~\ref{sec:backgrounds}).}
\centering
\ifthenelse{\boolean{cms@external}}{\resizebox{\columnwidth}{!}}{}
{
\begin{tabular}{ lcc }
Process & $\mu^{+}$& $\mu^{-}$ \\
\hline
Top quark (\ttbar~and tW) & $6837 \pm13$ & $6844 \pm 13$ \\
\wjets and \zjets & $2752 \pm 82 $ & $ 2487 \pm 76 $ \\
QCD multijet &$ \x\x308 \pm154$ & $ \x\x266 \pm 133$ \\
\hline
Single top quark $t$-channel&$1493 \pm 13 $ & $ \x948 \pm 10 $ \\
\hline
Total expected& $11390 \pm 175$ &$ 10545 \pm 154 $\\
\hline
Data&11877&  11017  \\
\end{tabular}
}
\label{tab:yield}
\end{table}
To analyze the kinematics of single top quark production, the momentum four-vectors of the top quarks are reconstructed from the decay products, muons, neutrinos, and b-jet candidates. The $\pt$ of the neutrino can be inferred from $\mptvec$. The longitudinal momentum of the neutrino, $p_{z,\nu}$, is inferred assuming energy-momentum conservation at the W$\mu\nu$ vertex and constraining the W boson mass to $m_{\PW}  = 80.4\GeV$~\cite{Agashe:2014kda}:
\begin{linenomath}
\begin{equation}
\label{eq:nusolver}
p_{z,\nu} =\frac{\Lambda p_{z,\mu}}{p_{\mathrm{T},\mu}^2}\pm\frac{1}{p_{\mathrm{T},\mu}^2}\sqrt{\Lambda^2 p_{z,\mu}^2-p_{\mathrm{T},\mu}^2(E_{\mu}^{2} \ETslash^2-\Lambda^2)},
\end{equation}
\end{linenomath}
where
\begin{linenomath}
\begin{equation}
\label{eq:lambda}
\Lambda=\frac{m_{\PW}^2}{2}+\vec{p}_{\mathrm{T},\mu}\cdot{ \mptvec },
\end{equation}
\end{linenomath}
and $E_{\mu}^{2}=p_{\mathrm{T},\mu}^2 + p_{z,\mu}^2$ denotes the muon energy.
In most of the cases this leads to two real solutions for $p_{z,\nu}$ and  the solution with the smallest absolute value is chosen~\cite{Abazov:2009ii,Aaltonen:2009jj}. For some events the discriminant in Eq.~(\ref{eq:nusolver}) becomes negative leading to complex solutions for $p_{z,\nu}$. In this case the imaginary component is eliminated by modification of $\mptvec$ so that $\mTW = \mW$, while still respecting the $\mW$ constraint. This is achieved by imposing that the determinant, and thus the square-root term in Eq.~(\ref{eq:nusolver}), is null. This condition gives a quadratic relation between $p_{x,\nu}$ and $p_{y,\nu}$ with two possible solutions, and one remaining degree of freedom. The solution is chosen by finding the neutrino transverse momentum $\vec{p}_{\mathrm{T},\nu}$  that has the minimum vectorial distance from the \mptvec in the $p_x\!-\!p_y$ plane.
The top quark candidate is reconstructed by combining the reconstructed W boson and the b-jet candidate. In the \threeJtwoT category, the b-jet candidate is the one with the higher b tagging discriminator value while the more central jet is used to reconstruct the top quark in the \twoJzeroT category.

\section{Background yields and modelling}
\label{sec:backgrounds}
\begin{figure*}[h]
\begin{center}
\includegraphics[width=0.30\textwidth]{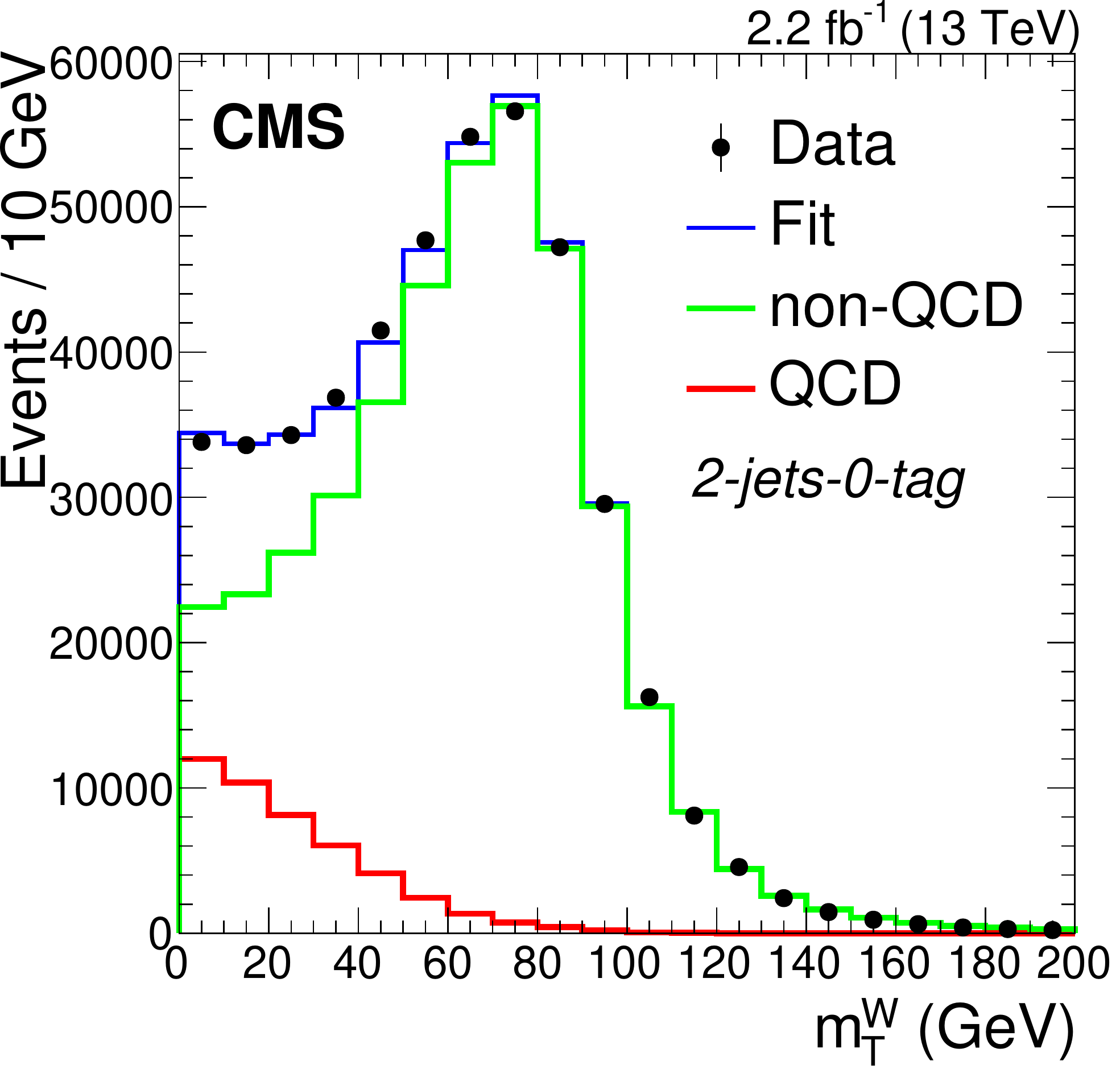}
\includegraphics[width=0.30\textwidth]{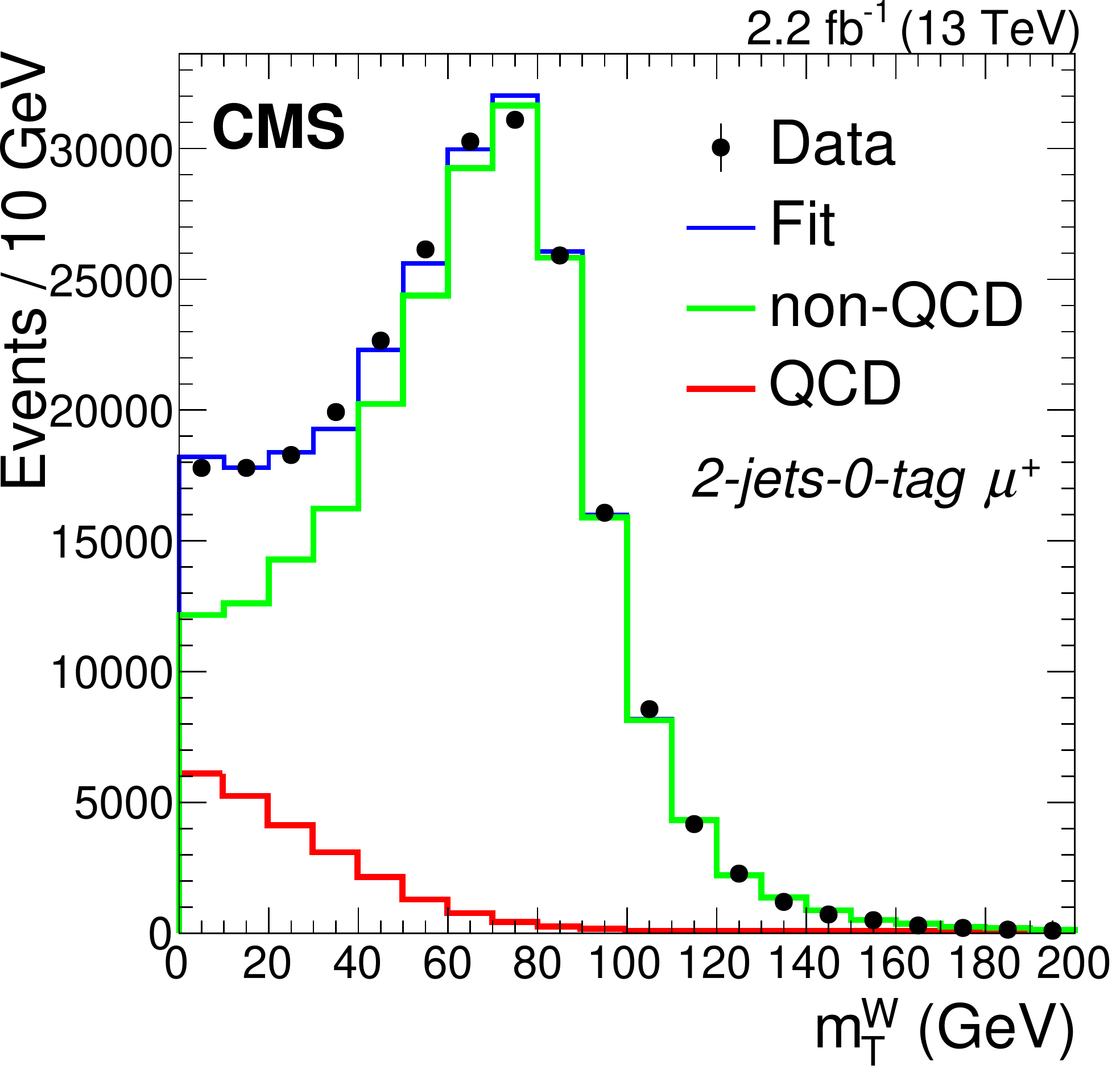}
\includegraphics[width=0.30\textwidth]{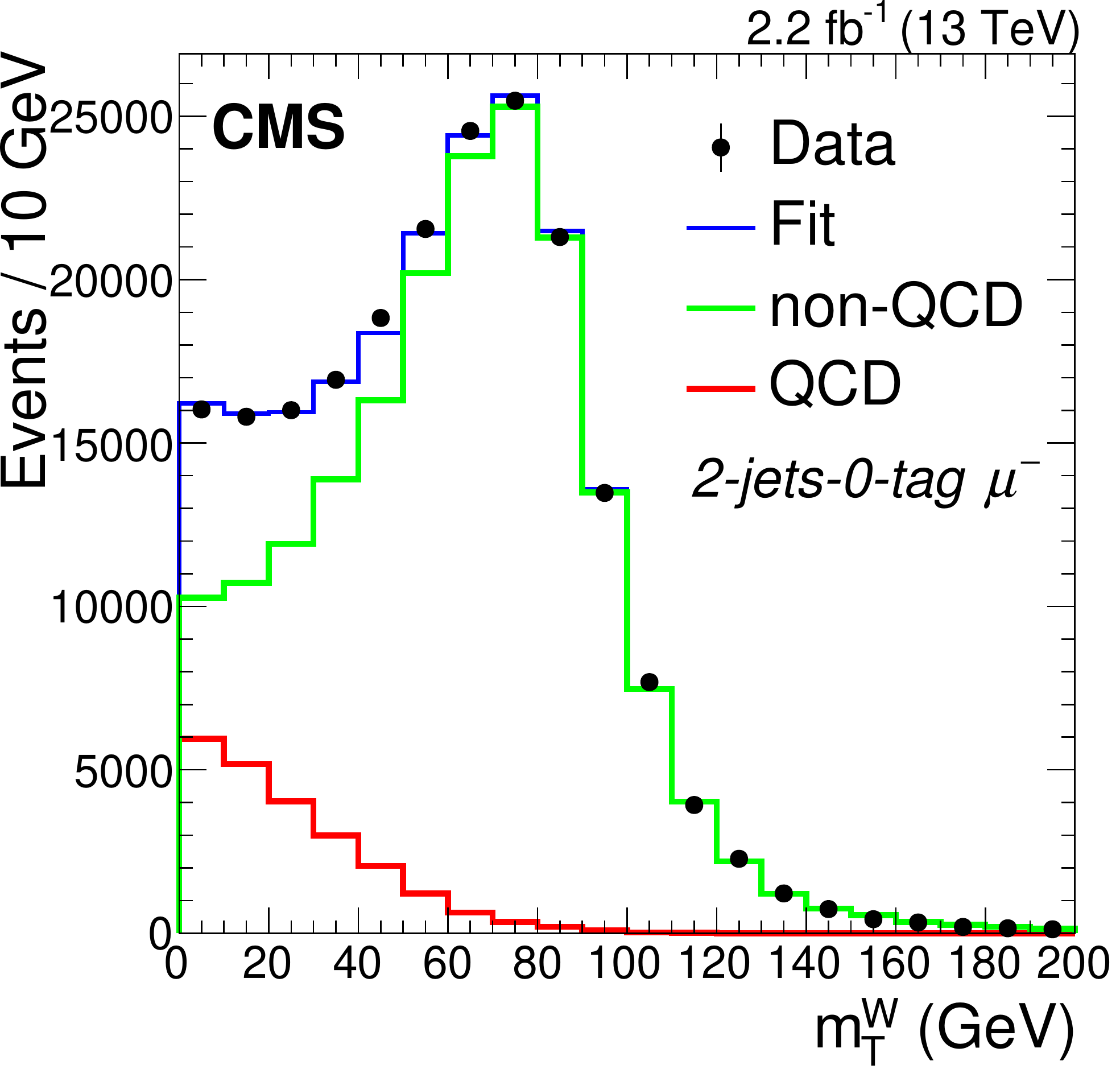}\\
\includegraphics[width=0.30\textwidth]{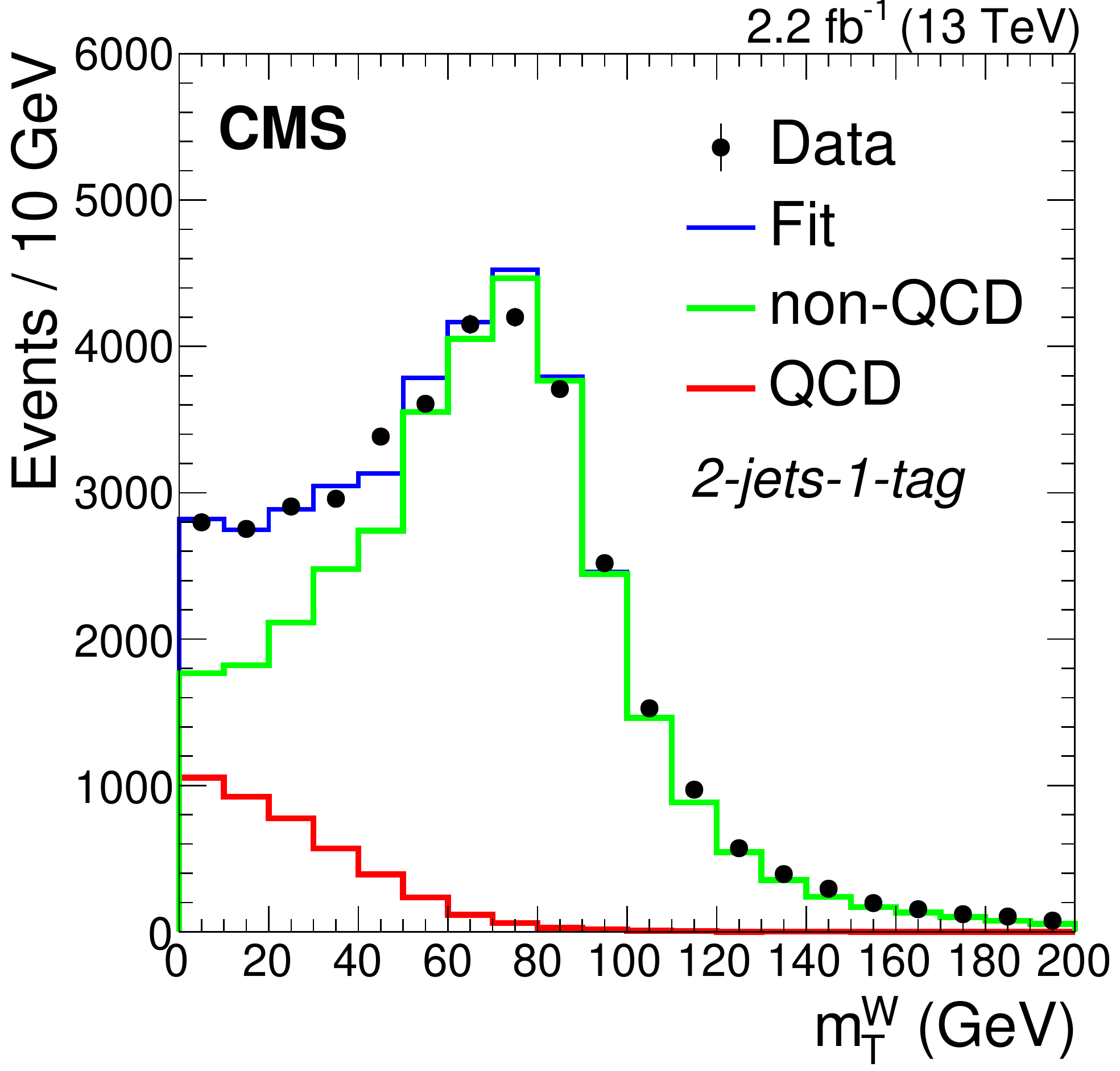}
\includegraphics[width=0.30\textwidth]{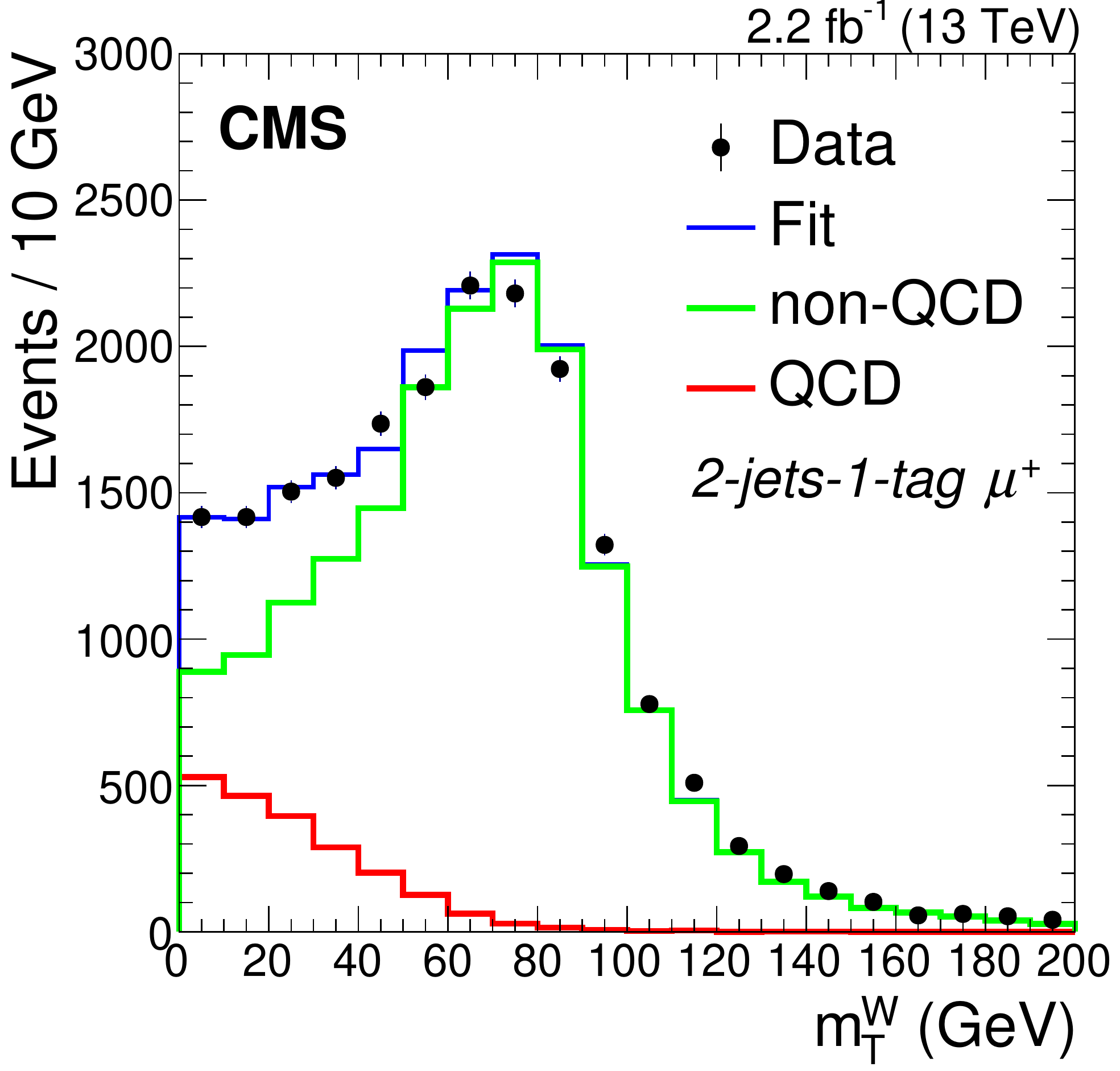}
\includegraphics[width=0.30\textwidth]{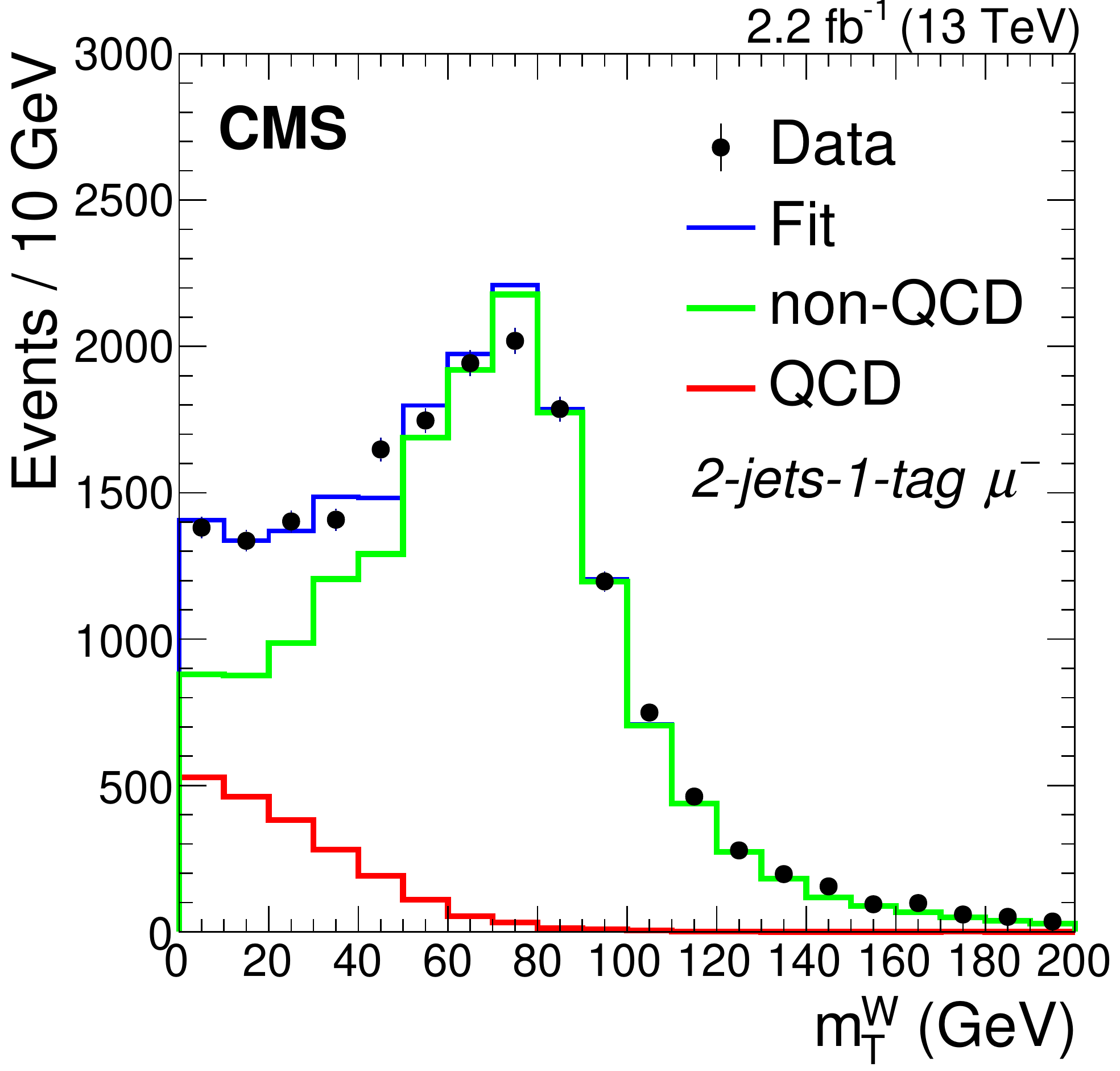}
\caption{\label{fig:QCD} Fit to the $\mTW$ distributions in the \twoJzeroT sample (upper row) and the \twoJoneT sample (lower row) for all events (left), for positively charged muons only (middle), and for negatively charged muons only (right). The QCD fit template is derived from a sideband region in data. Only statistical uncertainties are taken into account in the fit.}
\end{center}
\end{figure*}

The event yields for the various processes, summarized in
Table~\ref{tab:yield}, serve as the first order estimate of the respective
contributions to the data sample. The main background contributions come from \ttbar~production and the production of W bosons in association with jets. The validity of the MC simulation of these two processes is checked in data sideband regions enriched in these events. The modelling of the relevant kinematic variables for \ttbar~production can be checked in events with three jets, of which one or two are identified as stemming from b quark hadronization (\threeJoneT and \threeJtwoT), where \ttbar~events constitute by far the largest fraction of events. The \twoJzeroT region is enriched in W+jets events and is used to validate the modelling of the relevant variables for this background category. From these validations no indication of significant mismodelling of either \ttbar~production or the production of W bosons and jets is observed.
For the third important background category, QCD multijet production, reliable simulations are not available. The contribution from QCD multijet events is therefore suppressed as much as possible by requirements in the event selection and the remaining contamination is extracted directly from data. The $\mTW$ is well suited to effectively remove events arising from QCD multijet background as the shape of the distribution is different for QCD and non-QCD processes. In addition, the transverse mass is used to determine the remaining contribution of the QCD multijet background in the signal region. For this purpose, the requirement on $\mTW$ is removed and the entire $\mTW$ distribution is fitted using a maximum likelihood fit. The resulting yield of QCD multijet events is then extrapolated to the sample with $\mTW > 50\GeV$. Two probability distribution functions are used to fit the $\mTW$ distribution in data, one non-QCD distribution for all processes except the QCD multijet background, including $t$-channel signal, and one QCD distribution. For the former, the different non-QCD processes are added according to the MC-predicted contributions. The latter is extracted from a QCD-enriched data sample, defined by inverting the muon isolation requirement, with \PFrelIso $> 0.12$. The expected contamination from non-QCD processes in this region is around 10\%. Figure~\ref{fig:QCD} shows examples of the fitted $\mTW$ distributions in the most important region, the \twoJoneT signal region, inclusively and separately for events with positively and negatively charged muons. For these fits, only statistical uncertainties are taken into account. The validity of this procedure is tested on events in the \twoJzeroT category where the contribution of QCD multijet events is significantly larger than that of the \twoJoneT region (see also Fig.~\ref{fig:QCD}). When feeding the results of this QCD multijet background estimation into the procedure to extract the cross section of single top quark production, an uncertainty of 50\% is considered, which provides full coverage for all effects from variations in the rate and shape of this background contribution.

\section{Signal extraction strategy}
\label{sec:measurement}
To improve the discrimination between signal and background processes, an MVA technique is used to combine the discrimination power of several kinematic variables into one discriminator value. In this analysis, a total of 11 kinematic variables are combined into one single discriminator using the artificial neural network \textsc{NeuroBayes}~\cite{2006NIMPA.559..190F}, implemented in the TMVA~\cite{TMVA} package. The input variables are ranked according to their importance in Table~\ref{tab:mva_input}. The importance is defined as the loss of significance when removing this variable from the list. The variable with the largest discrimination power is the $\abs{\eta}$ of the light-quark jet. This importance is due to the fact that the presence of a light-quark jet in the forward direction is a typical feature of the topology of $t$-channel single top quark production. The second most important variable is the invariant mass of the reconstructed top quark, which discriminates processes with top quarks, from background processes without any produced top quark. All input variables are validated by comparing the distributions in data with those in the simulations. Simulated $t$-channel single top quark events are used as signal training sample, while simulated \ttbar~and \wjets~events, as well as QCD multijet events from a sideband region in data are used as background training samples, weighted according to their predicted relative contribution.
The neural network is trained on a subset of the simulated samples. Application on the remaining sample shows similar performance and no signs of overtraining are observed.
The neural network is trained in the inclusive \twoJoneT sample for events with positively and negatively charged muons, and afterwards applied to the \twoJoneT, \threeJoneT, and \threeJtwoT data samples, each further split in two, depending on the charge of the muon. In categories with ambiguity, the most forward jet is considered as the recoiling jet in the multivariate discriminator construction.

\begin{table*}[!h]
\begin{center}
\topcaption{\label{tab:mva_input} Input variables used in the neural network ranked according to their importance.}
\begin{tabular}{cll}
Rank & Variable & Description \\
\hline
\multirow{2}{*}{$1$} & \multirow{2}{*}{Light quark $|\eta|$} & Absolute value of the pseudorapidity of the \\
&                      & light-quark jet   \\
\hline
\multirow{2}{*}{$2$} & \multirow{2}{*}{Top quark mass} & Invariant mass of the top quark reconstructed  \\
&                & from muon, neutrino, and b-tagged jet     \\
\hline
3& Dijet mass & Invariant mass of the two selected jets        \\
\hline
4& Transverse W boson mass & Transverse mass of the W boson\\
\hline
5& Jet $\pt$ sum & Scalar sum of the transverse momenta of the two jets \\
\hline
\multirow{2}{*}{$6$} & \multirow{2}{*}{$\cos{\theta^*}$} & Cosine of the angle between the muon and the  \\
&                  & light-quark jet in the rest frame of the top quark       \\
\hline
\multirow{2}{*}{$7$} & \multirow{2}{*}{Hardest jet mass} & Invariant mass of the jet with the largest \\
&                 & transverse momentum    \\
\hline
\multirow{2}{*}{$8$} & \multirow{2}{*}{$\Delta R$ (light quark, b quark)} &   $\Delta R$ between the momentum vectors of the light-quark  \\
&                           & jet and the b-tagged jet.           \\
\hline
9& Light quark $\pt$ & Transverse momentum of the light-quark jet      \\
\hline
10& Light quark mass & Invariant mass of the light-quark jet         \\
\hline
\multirow{2}{*}{$11$}& \multirow{2}{*}{W boson $|\eta|$} &  Absolute value of the pseudorapidity of \\
& & the reconstructed W boson       \\
\hline
\end{tabular}
\end{center}
\end{table*}

\begin{figure*}[!h]
\begin{center}
\includegraphics[width=0.3\textwidth]{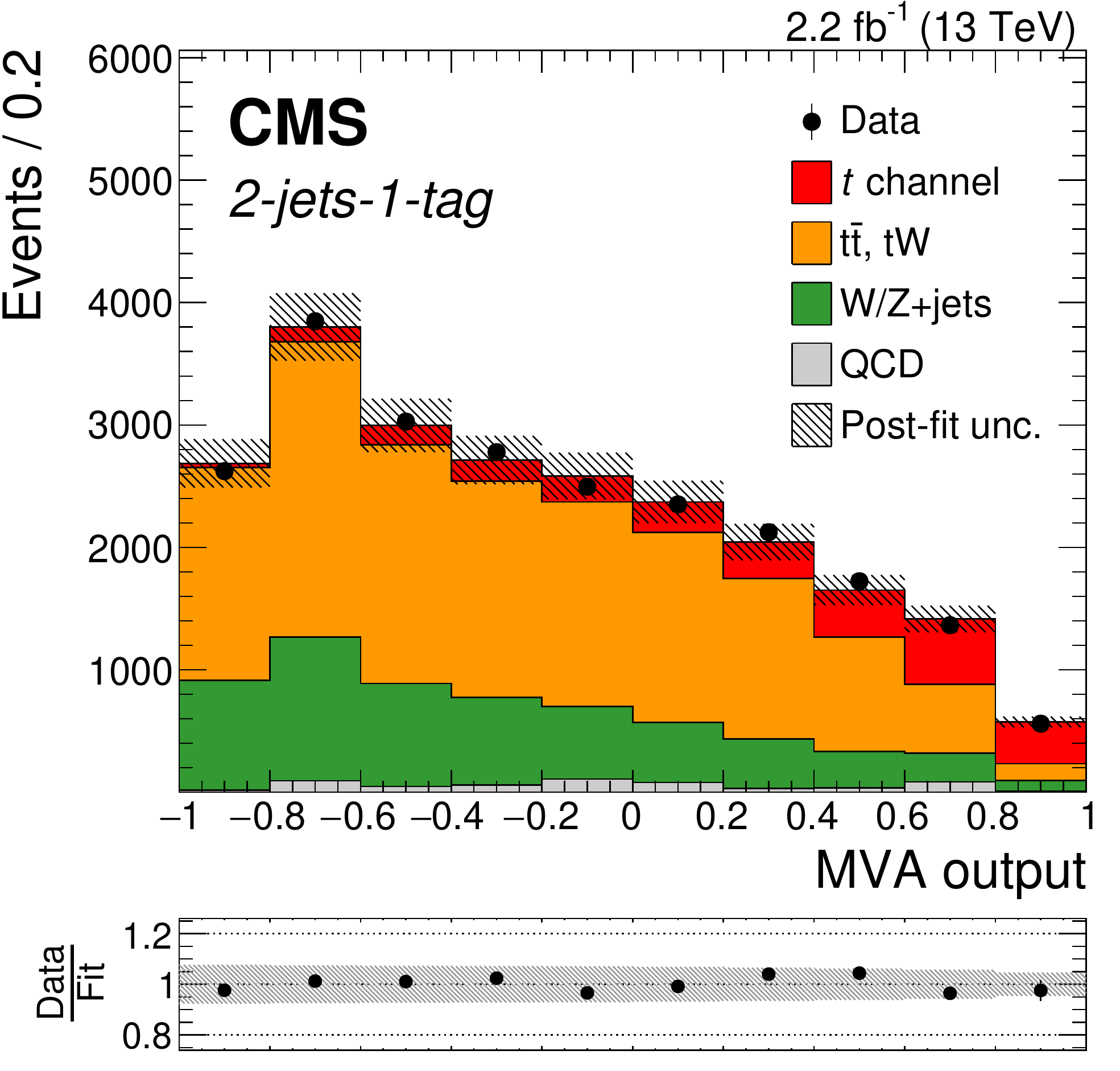}
\includegraphics[width=0.3\textwidth]{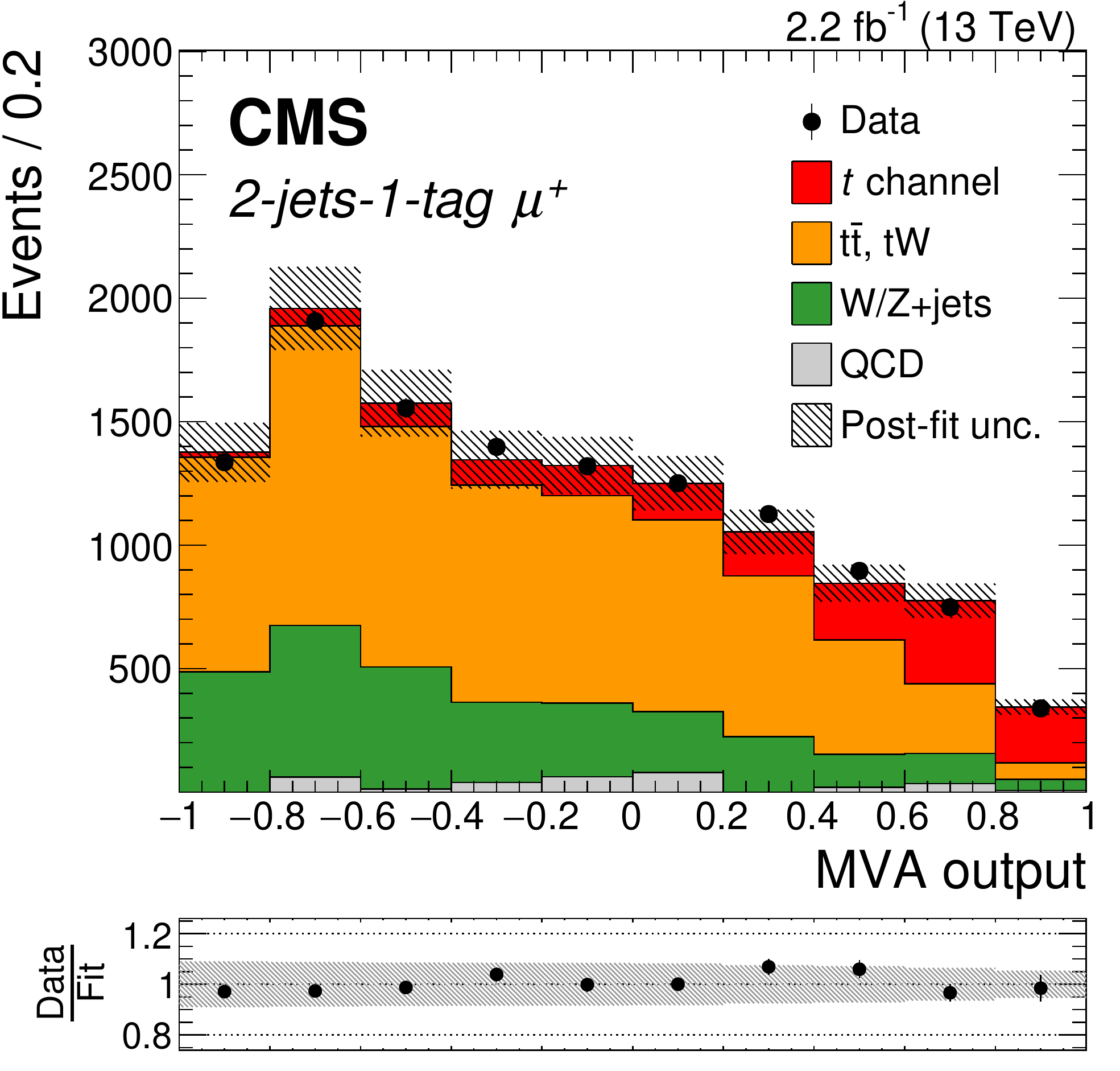}
\includegraphics[width=0.3\textwidth]{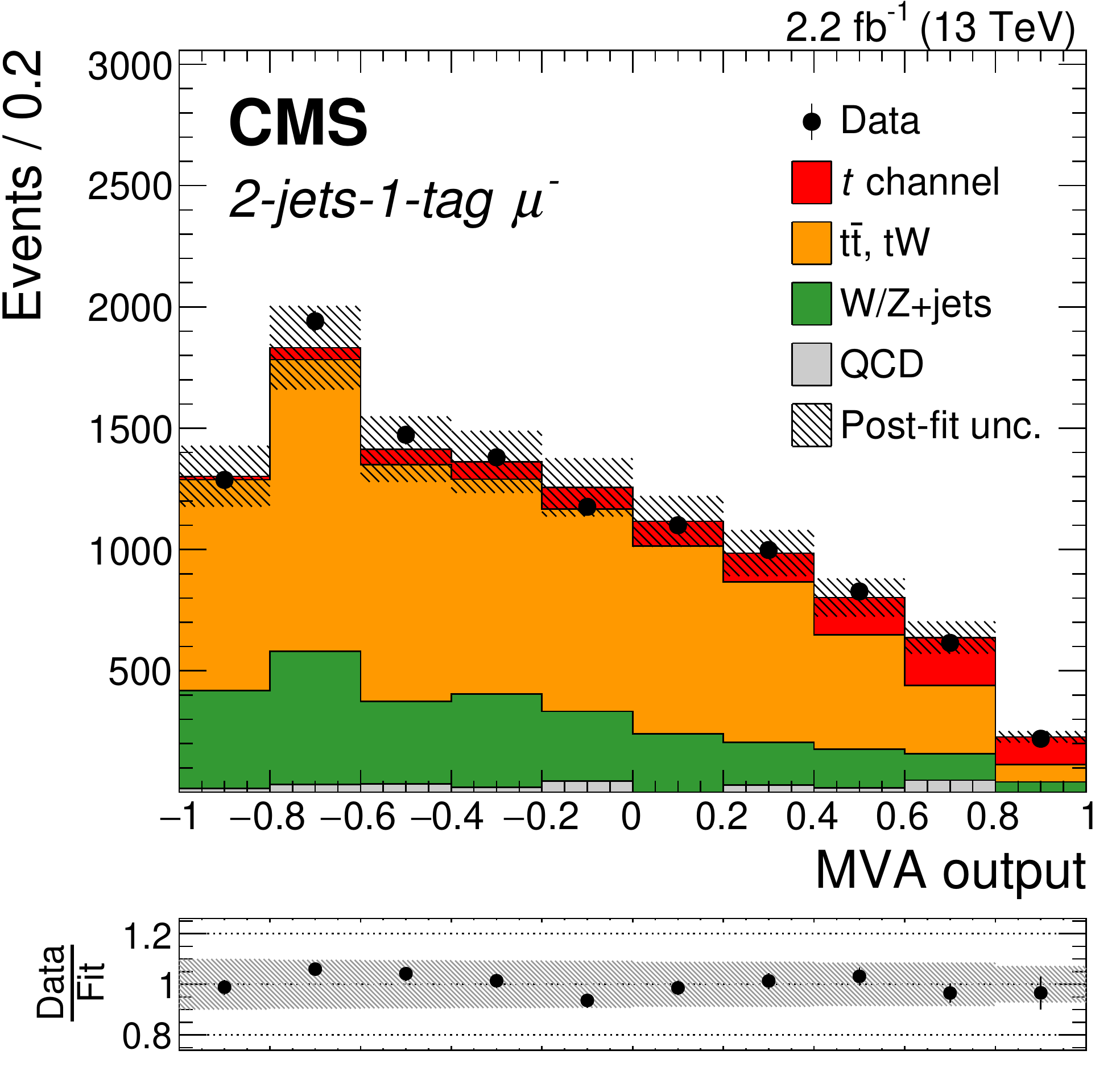}\\
\includegraphics[width=0.3\textwidth]{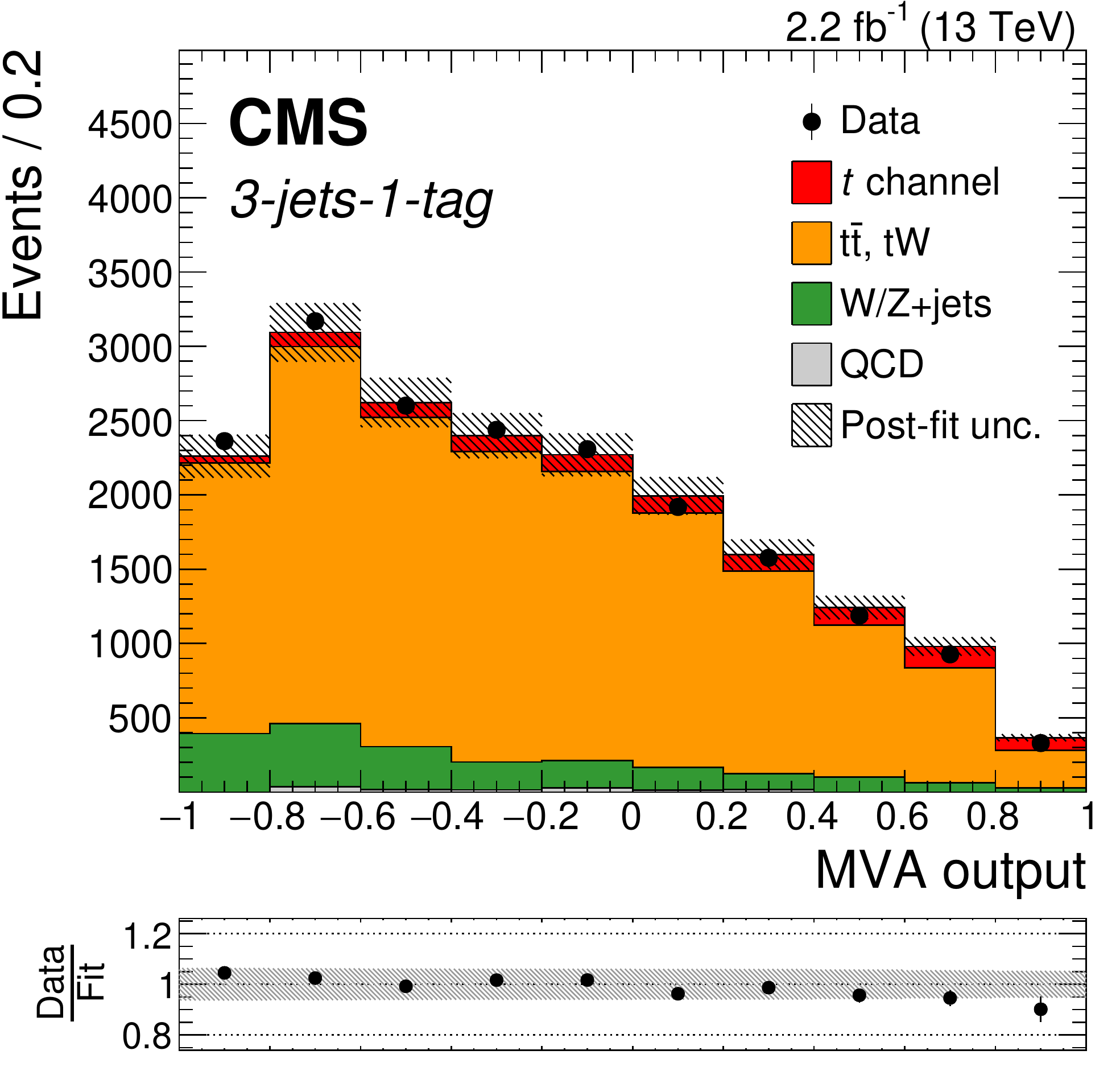}
\includegraphics[width=0.3\textwidth]{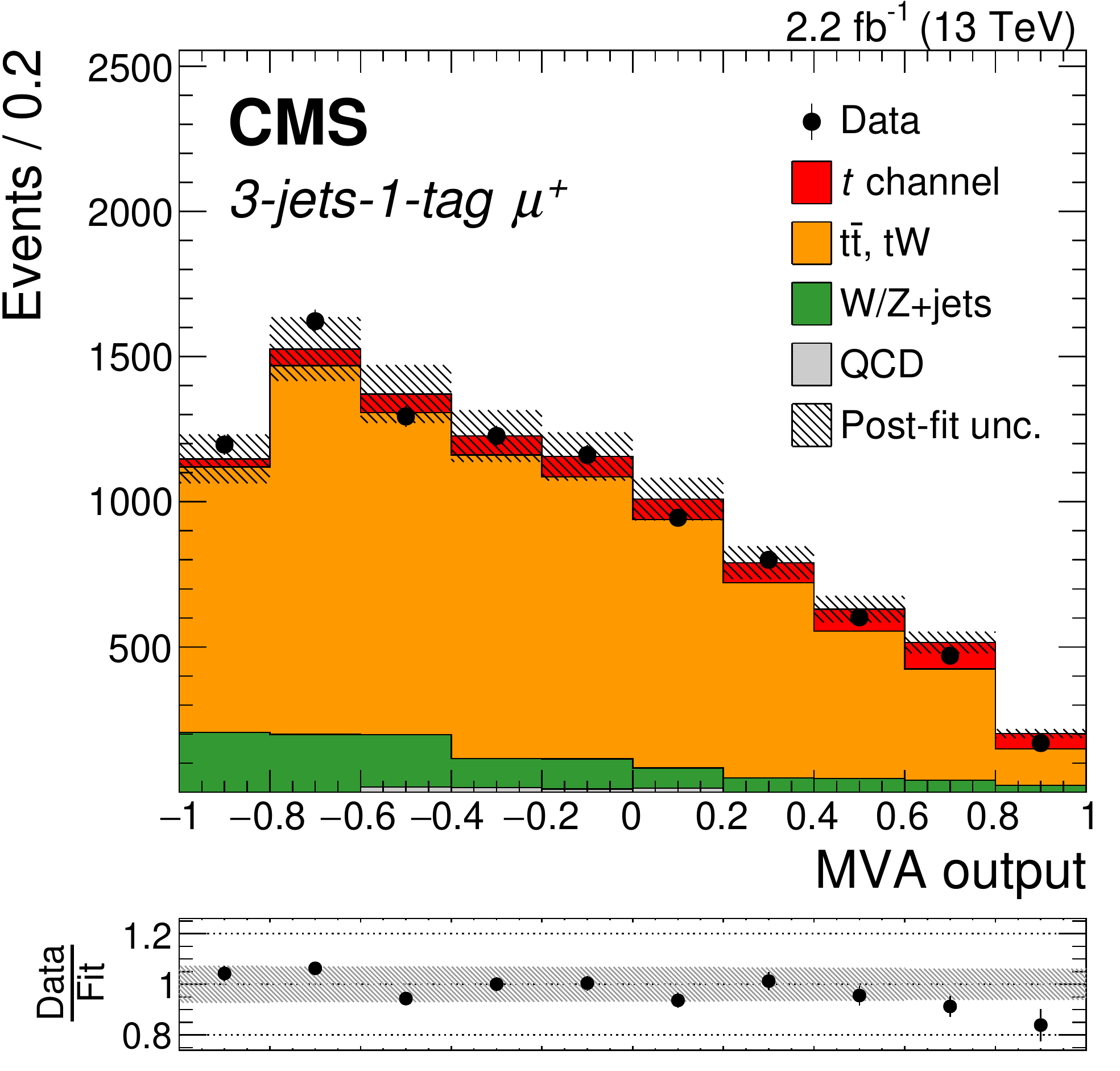}
\includegraphics[width=0.3\textwidth]{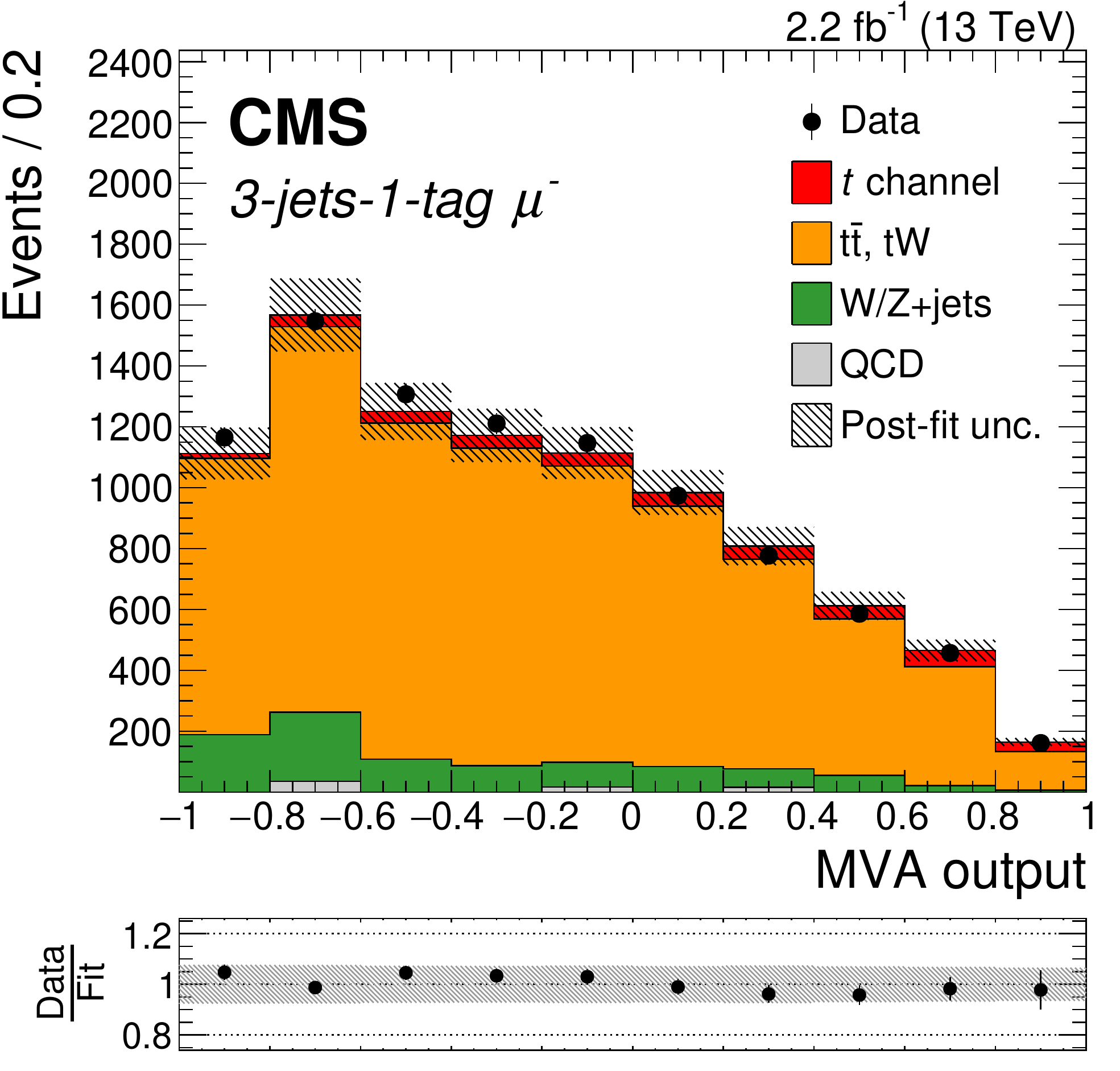}\\
\includegraphics[width=0.3\textwidth]{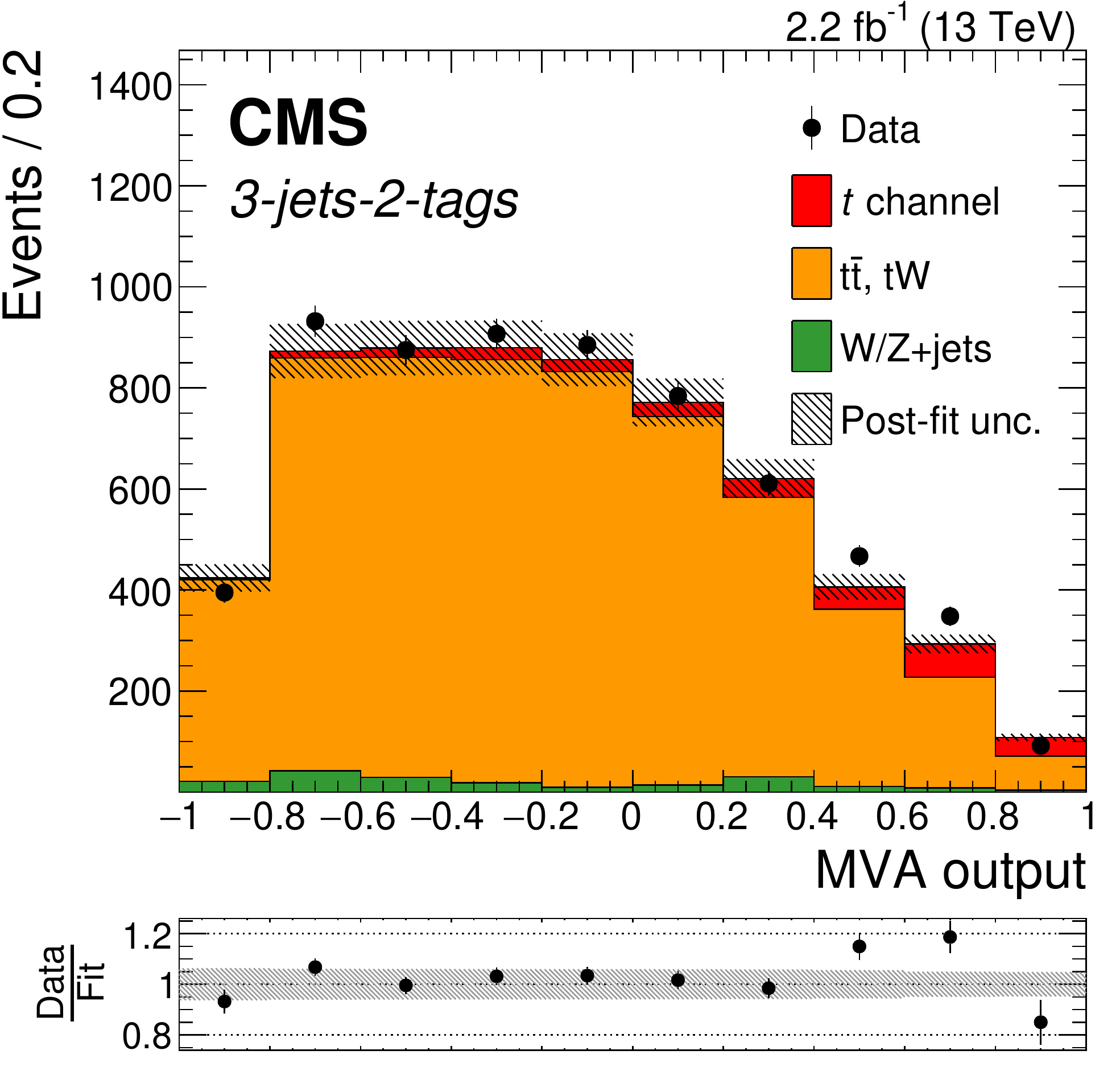}
\includegraphics[width=0.3\textwidth]{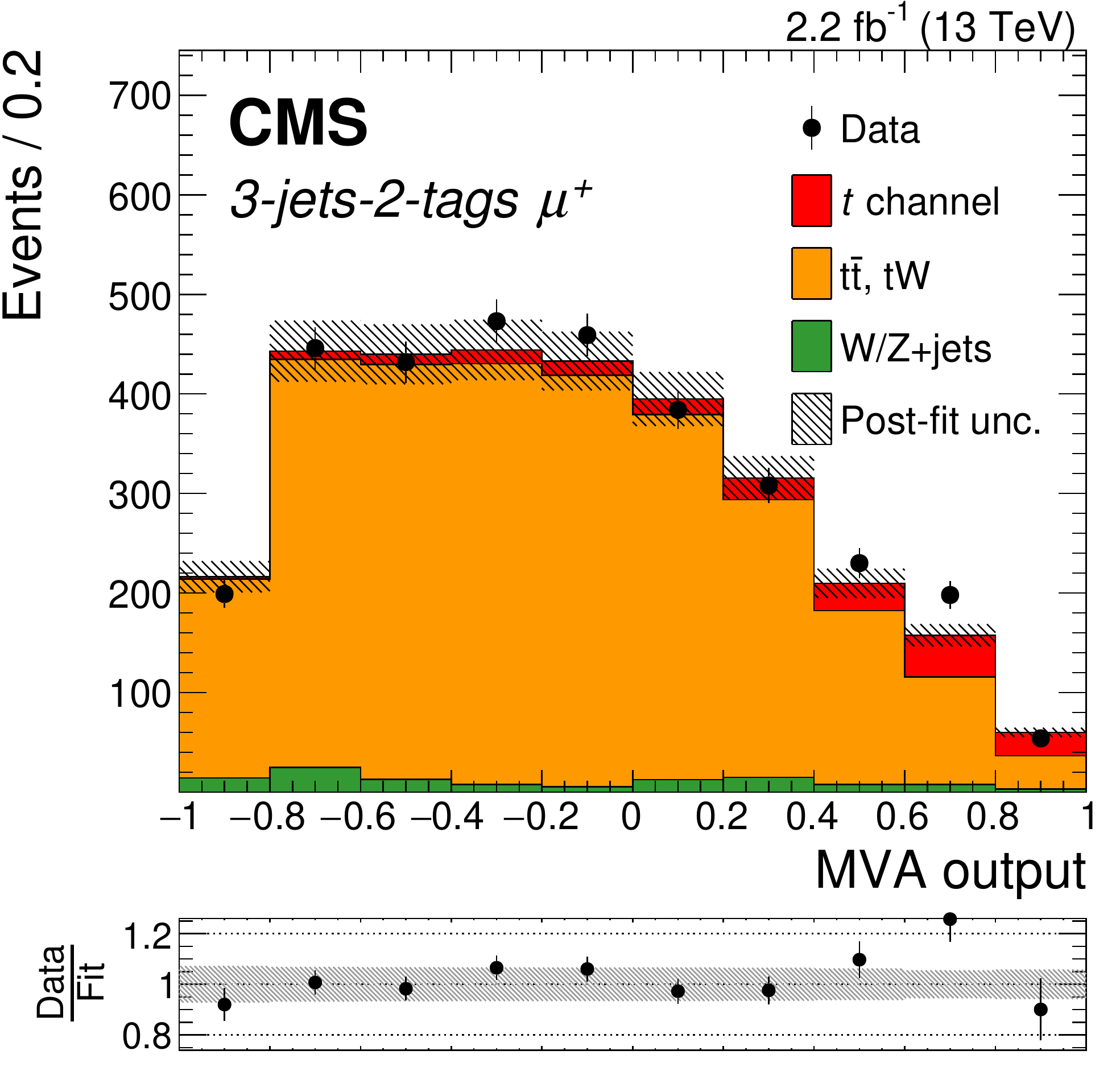}
\includegraphics[width=0.3\textwidth]{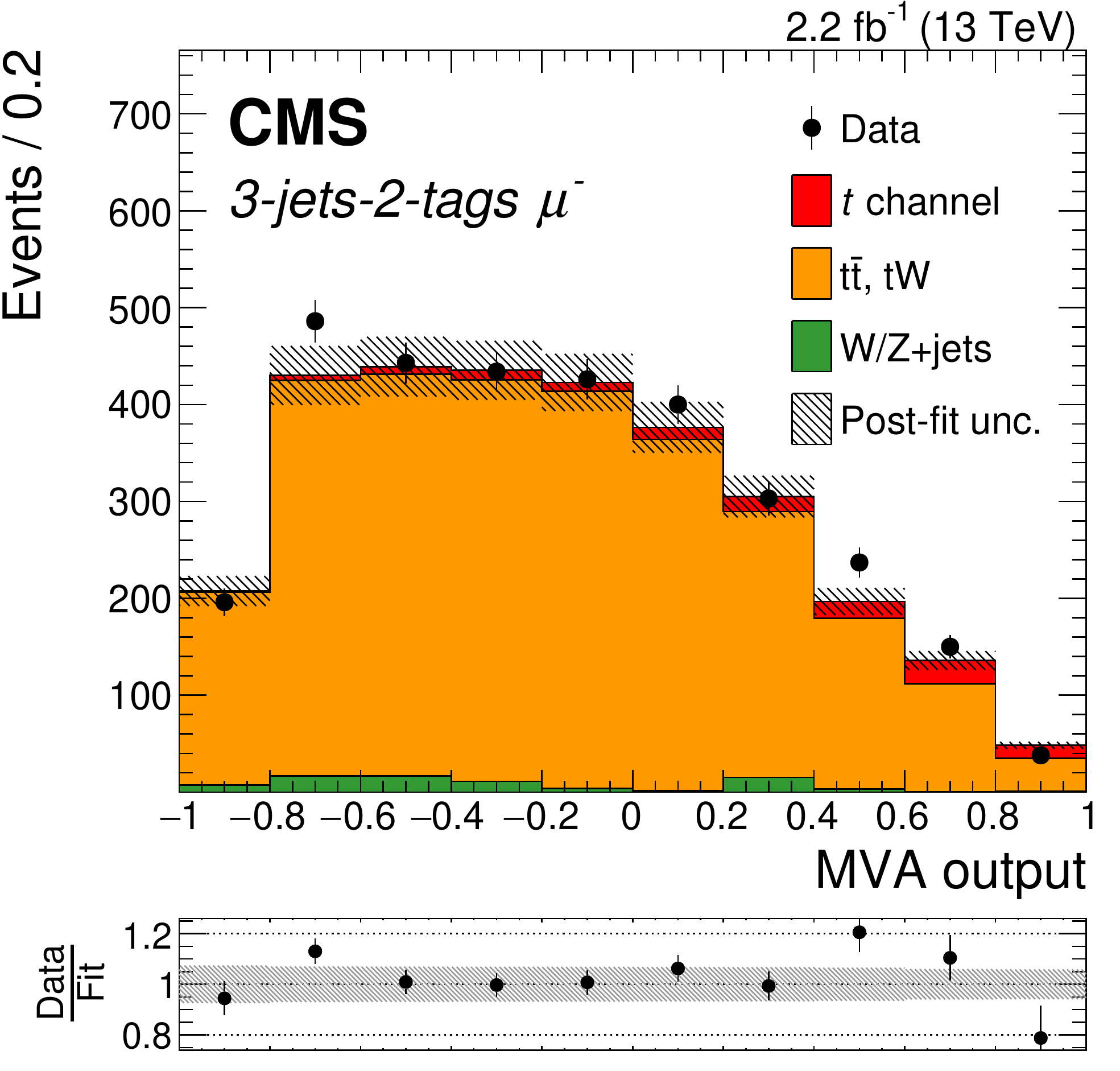}
\caption{\label{fit:result} Neural network distributions for all (left), positively (middle), and negatively (right) charged muons normalized to the yields obtained from the simultaneous fit in the \twoJoneT (upper), \threeJoneT (middle), and \threeJtwoT region (lower).
The ratio between data and simulated distributions after the fit is shown at the bottom of each figure.
The hatched areas indicate the post-fit uncertainties.}
\end{center}
\end{figure*}

To determine the signal cross sections, binned likelihood fits are performed on the distributions of the MVA discriminators. The background contributions are made up of three templates to account for: i) top quark production including \ttbar\,and tW, ii) electroweak production including W+jets and Z+jets processes, and iii) QCD multijet production.
The fit is performed using the Barlow--Beeston method~\cite{Barlow:1993dm} which correctly accounts for limited-size simulation samples.
The distributions of the MVA discriminators in the signal region (\twoJoneT) and the two control regions (\threeJoneT and \threeJtwoT) are fitted simultaneously. As the latter are dominated by \ttbar~events, including these control regions improves the precision of the \ttbar~contribution determination. The free parameters of the fit are the scale factor for the normalization of the single top quark production, the scale factors for the normalization of the background processes, and the ratio of single top quark to top antiquark production $R_{t\text{-ch.}}$.
The background scale factors are constrained by log-normal priors with an uncertainty of 10\% for the top quark background, 30\% for the electroweak background, and 50\% for the QCD multijet background. The latter is motivated by the uncertainties in the QCD estimation from data, while the other two are determined by the uncertainty on the theoretical cross sections.
The scale factors are defined as
\begin{linenomath}
\begin{equation}
S_i = \frac{N_i}{N_i ^{\text{pred.}}},
\label{eq:scalef}
\end{equation}
\end{linenomath}
where $N_i$ is the number of events after the fit, $N_i ^{\text{pred.}}$ the predicted number of events and $i$ the process category. Table~\ref{tab:fitresult} shows the results obtained from the fit for events with a positively charged muon. The fitted distributions are shown in Fig.~\ref{fit:result}.

\begin{table}[!h]
\begin{center}
\topcaption{\label{tab:fitresult} Scale factors from the fit for the normalization of events with a positively charged muon for the signal process, the background categories, and the ratio of single top quark to top antiquark production. The uncertainties include the statistical uncertainty and the experimental sources of uncertainty which are considered as nuisance parameters in the fit.}
\begin{tabular}{l|c}
Process  & Scale factor \\
\hline
Signal, $t$ channel & $1.13 \pm  0.08 $ \\
Top quark background ($\ttbar$ and tW) & $1.00 \pm  0.02$  \\
\wjets and \zjets & $1.11 \pm  0.09$  \\
QCD multijet  & $0.86 \pm  0.29$  \\
$R_{t\text{-ch.}}$  & $1.81 \pm  0.19$ \\
\end{tabular}
\end{center}
\end{table}

\section{Systematic uncertainties}
\label{sec:systematics}
The measurement of the cross section is affected by various sources of systematic uncertainties, which can be grouped into two categories, experimental uncertainties and theoretical uncertainties. Several of the former category of uncertainties are considered as nuisance parameters in the fit to the MVA discriminator distribution and are thus included in the total uncertainty of the fit. To determine the impact of the sources of the remaining uncertainties, pseudo-experiments are performed. Pseudo-data are drawn from the nominal samples. Fits to the discriminator distributions are performed with templates, including the variations in the shapes that correspond to systematic variations of one standard deviation. The difference between the mean values of the results from these fits, and from fits using the nominal shapes as fit templates, is taken as an estimation for the corresponding uncertainty. The contributions from different sources are summed together with the method in Ref.~\cite{Barlow:2003sg}: the asymmetric components of each uncertainty are treated as the standard deviations of two halved Gaussian functions, and thus the convolution of the resulting distributions for all uncertainties is performed by making use of Thi\'ele's semi-invariants.
\par
\textit{Experimental uncertainties --- included in the fit}
\par
The following sources of systematic uncertainty are included in the fit either
through the applied Barlow-Beeston method or by using nuisance parameters in
the fit (profiled uncertainties). By variations of the default samples, two dedicated templates corresponding to ${\pm}1$ standard deviations of the respective uncertainty source are created. The fit interpolates between these templates according to the actual value of the nuisance parameter.
\begin{itemize}
\item \textbf{Limited size of samples of simulated events}: To account for the
limited number of available simulated events the fit is performed using the
Barlow--Beeston method, and the effect is therefore
included in the total uncertainty of the fit. To estimate the impact of the
sample size the nominal central value is compared with the central value obtained without the
Barlow--Beeston method. The latter effectively corresponds to
assuming an infinite size of the samples of simulated events.
\item \textbf{Jet energy scale (JES)}: All reconstructed jet four-momenta in simulated events are simultaneously varied according to the $\eta$- and $\pt$-dependent uncertainties in the JES~\cite{Chatrchyan:2011ds}. This variation in jet four-momenta is also propagated to $\PTslash$.
\item \textbf{Jet energy resolution (JER)}: A smearing is applied
to account for the difference in the JER between simulation and data~\cite{Chatrchyan:2011ds}, increasing
or decreasing the resolutions by their uncertainties.
\item  \textbf{The b tagging}: b tagging and misidentification efficiencies are estimated from control samples in 13\TeV data~\cite{CMS-PAS-BTV-15-001}. Scale factors are applied to the simulated samples to reproduce efficiencies observed in data and the corresponding uncertainties are propagated as systematic uncertainties.
\item \textbf{Muon trigger and reconstruction}: Single-muon trigger efficiency and reconstruction efficiency are estimated with a ``tag-and-probe'' method~\cite{Khachatryan:2010xn} from Drell--Yan events in the Z boson mass peak.
To take the difference in kinematic properties between Drell--Yan and the single top quark process into account, an additional systematic uncertainty depending on the number of jets in an event is applied.
\end{itemize}
\par
\textit{Experimental uncertainties --- not included in the fit}
\par
\begin{itemize}
\item \textbf{Pileup}: The uncertainty in the average expected number of
pileup interactions is propagated as a source of systematic uncertainty to
this measurement by varying the minimum bias cross section by $\pm$5\%. The
effect on the result is found to be negligible and is therefore not
considered further.
\item \textbf{Luminosity}: The integrated luminosity is known with a relative
uncertainty of  ${\pm}\lumiunc$~\cite{lumi2015}.
\end{itemize}
\par
\textit{Theoretical uncertainties}
\par
\begin{itemize}
\item \textbf{Signal modelling}: To estimate the influence of possible mismodelling of the signal process, the default sample (MG5\_a\MCATNLO) is compared to a sample generated with \POWHEG, another NLO matrix-element generator. The effect of different PS models is estimated by comparing the default sample (MG5\_a\MCATNLO interfaced with \PYTHIA) with a sample using a different PS description (MG5\_a\MCATNLO interfaced to \HERWIGpp).
\item \textbf{\boldmath$\bbbar$~modelling}: For the estimation of the uncertainty due to possible mismodelling of the \ttbar~background, the same procedure as for the signal modelling is applied. The default sample, generated with \POWHEG, is compared to a sample generated with MG5\_a\MCATNLO to estimate the impact of the choice of the matrix-element generator, and the two PS models implemented in \PYTHIA and \HERWIGpp~\cite{Bahr:2008pv} are compared to estimate the influence of the PS modelling.
\item \textbf{W+jets modelling}: The impact of incorrectly modelled relative fractions of W boson production in association with heavy flavour jets in the W+jets sample is estimated by varying the fractions of W+b and W+c events independently by ${\pm}30\%$.
\item \textbf{Modelling of the top quark \boldmath$\pt$}: Differential measurements of the top quark $\pt$ in \ttbar~events~\cite{Khachatryan:2015oqa} have shown that a harder spectrum is predicted than observed. Therefore the results derived using the default simulation for \ttbar~are compared to the results using simulated \ttbar~events that are reweighted according to the observed difference between data and simulation in Ref.~\cite{Khachatryan:2015oqa}
\item \textbf{Renormalization and factorization scale uncertainty (\boldmath$\mu_{\mathrm{R}}/\mu_{\mathrm{F}}$)}: The uncertainties due to variations in the renormalization and factorization scales are studied for the signal process, tW, $\ttbar$, and $\wjets$ by reweighting the distributions with different combinations of halved/doubled factorization and renormalization scales. The effect is estimated for each process separately.
\item \textbf{PDF}: The uncertainty due to the choice of PDFs is estimated using reweighted histograms derived from all PDF sets of NNPDF~3.0~\cite{Botje:2011sn}.
\end{itemize}

Different contributions to the uncertainty on cross sections are summarised in Table~\ref{tab:systematics}.
Several of the experimental sources of uncertainty are treated as nuisance
parameters in the fit which results in a single uncertainty of the fit
including also the statistical contribution. By fixing all nuisance parameters
the statistical uncertainty can be obtained, including the uncertainty
due to the size of the samples of simulated events. The contribution due to
the profiled experimental uncertainties is derived by subtracting the statistical term quadratically from the fit uncertainty.
The breakdown of sources of uncertainty that are included in the fit, listed
in Table~\ref{tab:systematics_exp}, is for illustration only. The estimates of
the profiled systematic uncertainties are obtained by comparing the
uncertainty of the fit including all nuisance parameters with the uncertainty
of the fit where one source of uncertainty is kept fixed while all others are
included via nuisance parameters. The impact of the size of the samples of
simulated events is estimated as described above.

\begin{table*}
\topcaption{Relative impact of systematic uncertainties with respect to the observed cross sections as well as the top quark to top antiquark cross section ratio.
Uncertainties are grouped and summed together with the method suggested in Ref.~\cite{Barlow:2003sg}.}
\centering
\ifthenelse{\boolean{cms@external}}{}{\resizebox{\columnwidth}{!}}
{
\begin{tabular}{ l|cccc }
Uncertainty source  & $\Delta \sigma_{t{\text{-ch.}}, \PQt+\PAQt}/\sigma_{t{\text{-ch.}}, \PQt+\PAQt}^{\text{obs}}$ & $\Delta \sigma_{t{\text{-ch.}}, \PQt}/\sigma_{t{\text{-ch.}}, \PQt}^{\text{obs}}$  & $\Delta \sigma_{t{\text{-ch.}},\PAQt}/\sigma_{t{\text{-ch.}},\PAQt}^{\text{obs}}$ & $\Delta R_{t\text{-ch.}} / R_{t\text{-ch.}}$\\
\hline
Statistical uncert. & $\pm$5.5\% & $\pm$5.3\%  & $\pm$11.5\% & $\pm$9.7\%\\
Profiled exp. uncert. & $\pm$5.2\%  & $\pm$5.7\% & $\pm$4.9\%  & $\pm$3.3\%\\
\hline
Total fit uncert. & $\pm$7.6\% & $\pm$7.8\% & $\pm$12.5\% & $\pm$10.3\%\\
\hline
Integrated luminosity & $\pm$2.3\% & $\pm$2.3\%  & $\pm$2.3\% & ---\\
\hline
Signal modelling & $\pm$6.9\% & $\pm$8.2\%  & $\pm$8.5\% & $\pm$5.3\%\\
\ttbar~modelling & $\pm$3.9\% & $\pm$4.3\%  & $\pm$4.5\% & $\pm$4.0\%\\
\wjets modelling & $-$1.8/$+$2.1\% & $-$1.6/$+$2.3\% & $-$2.5/$+$2.3\%& $-$1.7/$+$2.0\%\\
$\mu_{\mathrm{R}}/\mu_{\mathrm{F}}$ scale $t$-channel & $-$4.6/$+$6.1\% & $-$5.7/$+$5.2\%  & $-$7.2/$+$5.1\%& $-$0.7/$+$1.2\%\\
$\mu_{\mathrm{R}}/\mu_{\mathrm{F}}$ scale \ttbar & $-$3.5/$+$2.9\% & $-$3.5/$+$4.1\%  & $-$4.7/$+$3.1\%& $-$1.1/$+$1.0\%\\
$\mu_{\mathrm{R}}/\mu_{\mathrm{F}}$ scale tW & $-$0.3/$+$0.5\% & $-$0.6/$+$0.8\%  & $-$1.1/$+$0.7\%& $-$0.2/$+$0.1\%\\
$\mu_{\mathrm{R}}/\mu_{\mathrm{F}}$ scale \wjets & $-$2.9/$+$3.7\% & $-$3.5/$+$3.0\%  & $-$4.9/$+$3.8\%& $-$1.2/$+$0.9\%\\
PDF uncert.  & $-$1.5/$+$1.9\% & $-$2.1/$+$1.6\%  & $-$1.8/$+$2.1\%  & $-$2.2/$+$2.5\%\\
Top quark $\pt$ modelling & $\pm$0.1\% & $\pm$0.2\%  & $\pm$0.2\% & $\pm$0.1\%\\
\hline
Total theory uncert. & $-$10.7/$+$11.1\% & $-$12.2/$+$12.1\% & $-$13.6/$+$12.9\%  & $\pm$7.5\%\\
\hline
Total uncert. & $-$13.4/$+$13.7\% & $\pm$14.7\%  & $-$18.7/$+$18.2\%  & $\pm$12.7\%\\
\end{tabular}
}
\label{tab:systematics}
\end{table*}

\begin{table*}
\topcaption{Relative impact of the experimental systematic uncertainties
included in the fit with respect to the observed cross sections as well as the top quark to top antiquark cross section ratio.
The impact due to the size of the samples of simulated events is estimated by
comparing the central values obtained by applying or not applying the
Barlow--Beeston method in the fit. All other estimates are obtained by fixing one uncertainty at a time and considering all others as nuisance parameters in the fit and comparing to the uncertainty obtained when treating all uncertainty sources as nuisance parameters. These numbers are for illustration only, the uncertainty quoted for the result is the total experimental uncertainty from the fit.}
\centering
\begin{tabular}{ l|cccc }
Uncertainty source  & $\Delta \sigma_{t{\text{-ch.}}, \PQt+\PAQt}/\sigma_{t{\text{-ch.}}, \PQt+\PAQt}^{\text{obs}}$ & $\Delta \sigma_{t{\text{-ch.}}, \PQt}/\sigma_{t{\text{-ch.}}, \PQt}^{\text{obs}}$  & $\Delta \sigma_{t{\text{-ch.}},\PAQt}/\sigma_{t{\text{-ch.}},\PAQt}^{\text{obs}}$ & $\Delta R_{t\text{-ch.}} / R_{t\text{-ch.}}$\\
\hline
MC samples size & $\pm$3.4\% & $\pm$4.1\% & $\pm$3.8\% & $\pm$3.2\% \\
JES & $\pm$4.1\% &$\pm$4.7\% & $\pm$3.5\% & $\pm$2.1\%\\
JER & $\pm$1.7\% & $\pm$1.2\% & $\pm$2.4\% & $\pm$0.6\%\\
b tagging efficiency & $\pm$1.9\% &$\pm$2.0\% & $\pm$1.8\% & $\pm$1.4\%\\
Mistag probability & $\pm$0.9\% & $\pm$0.6\% & $\pm$0.8\% & $\pm$0.5\%\\
Muon reco./trigger & $\pm$2.0\% & $\pm$2.3\%  & $\pm$1.9\% & $\pm$1.8\%\\
\end{tabular}
\label{tab:systematics_exp}
\end{table*}

\section{Results}
\label{sec:results}
\begin{figure}[!t]
\begin{center}
\includegraphics[width=\cmsFigWidth]{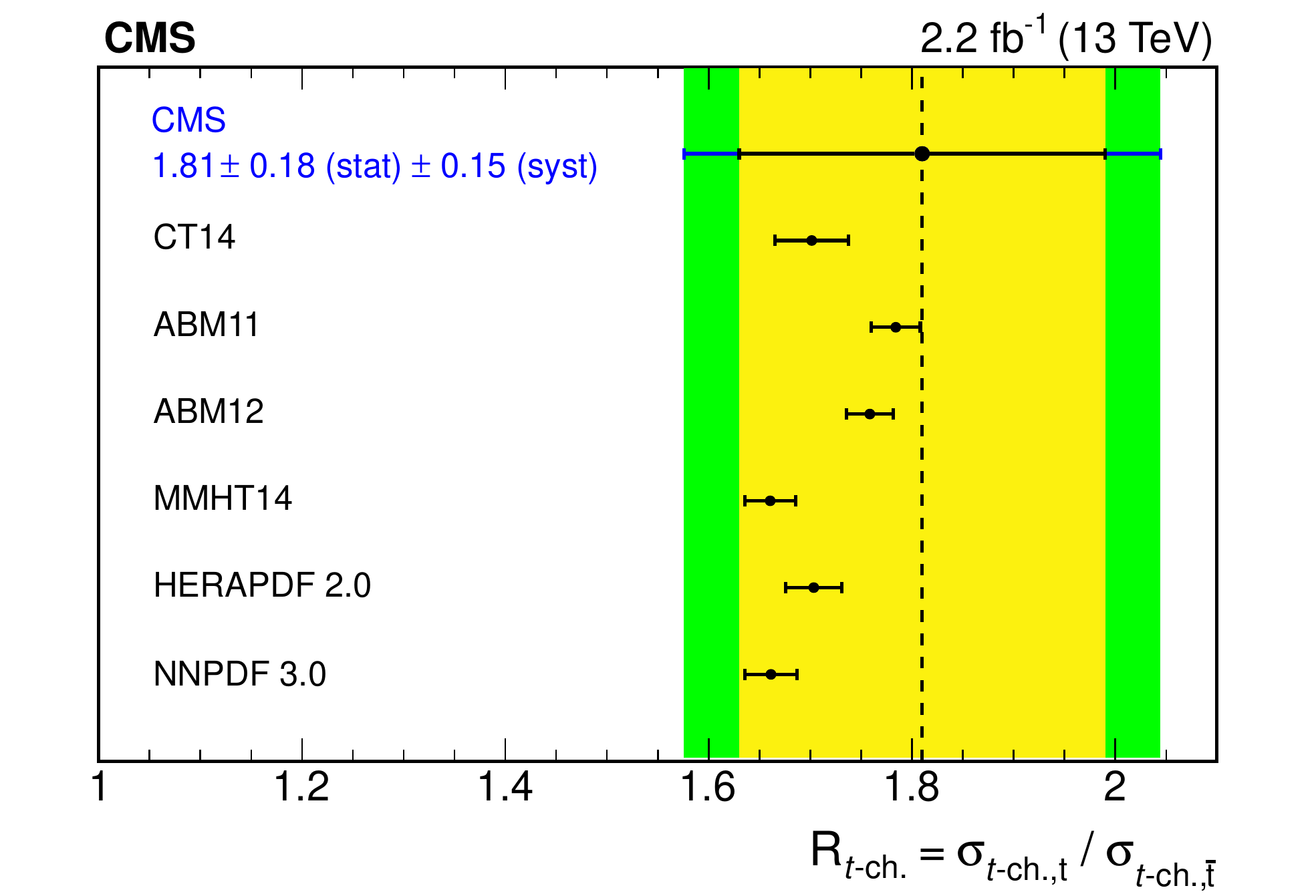}
\caption{\label{fig:ratioplot} Comparison of the measured $R_{t\text{-ch.}}$ (dotted line) with the prediction from different PDF sets: CT14 NLO~\cite{CT14}, ABM11 NLO and ABM12 NNLO~\cite{ABM}, MMHT14 NLO~\cite{MMHT14}, HERAPDF2.0 NLO~\cite{HERAPDF}, NNPDF 3.0 NLO~\cite{NNPDF}. The {\sc PowHeg} 4FS calculation is used. The nominal value for the top quark mass is $172.5\GeV$. The error bars for the different PDF sets include the statistical uncertainty, the uncertainty due to the factorization and renormalization scales, derived varying both of them by a factor 0.5 and 2, and the uncertainty in the top quark mass, derived varying the top quark mass between 171.5 and 173.5$\GeV$. For the measurement, the inner and outer error bars correspond to the statistical and total uncertainties, respectively.}
\end{center}
\end{figure}

\begin{figure}[!h]
\begin{center}
\includegraphics[width=\cmsFigWidth]{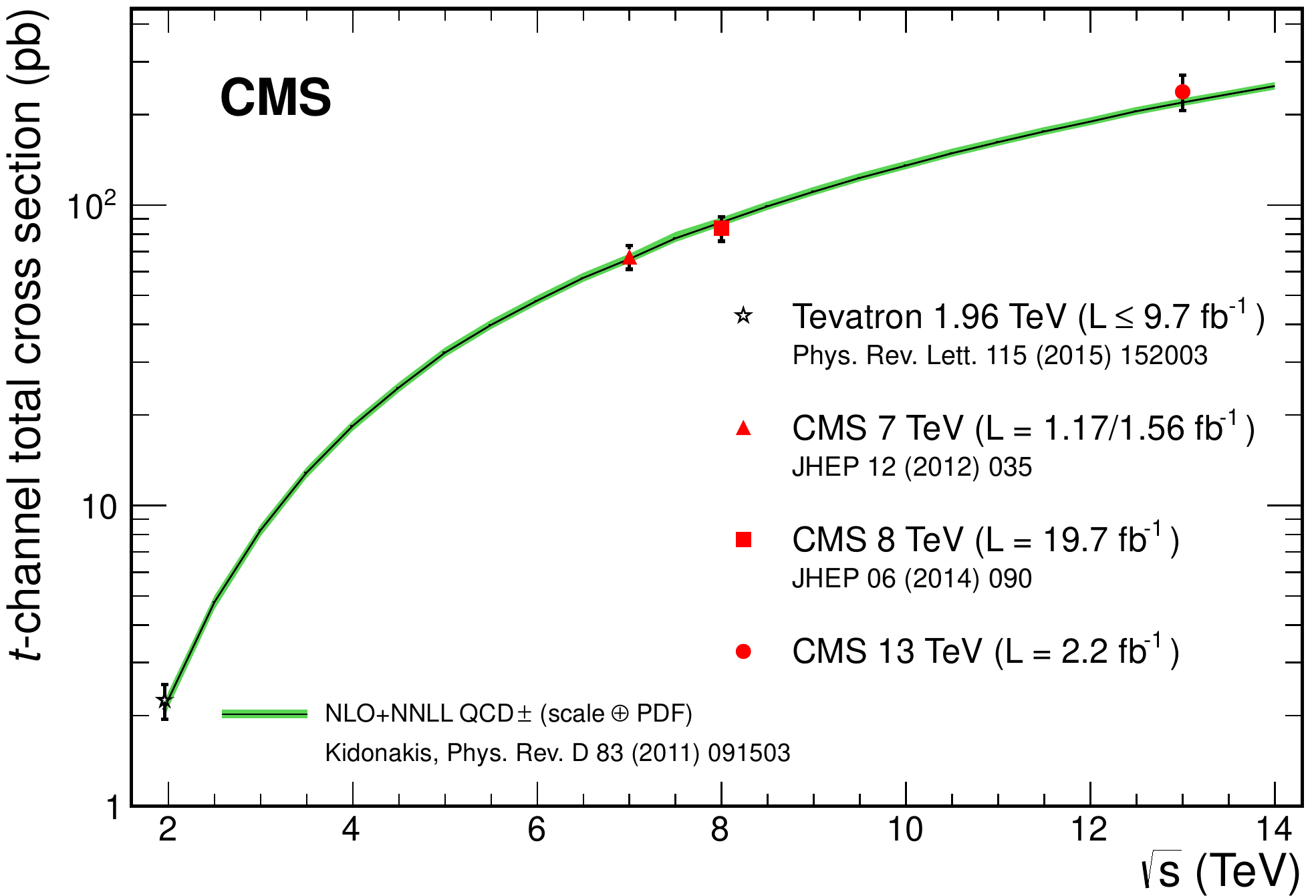}
\caption{\label{fig:finalplot} The summary of the most precise CMS measurements~\cite{Chatrchyan:2012ep,PASSingleTopCrossSection} for the total $t$-channel single top quark cross section, in comparison with NLO+NNLL QCD calculations~\cite{Kidonakis:2011wy}. The combination of the Tevatron measurements~\cite{PhysRevLett.115.152003} is also shown.}
\end{center}
\end{figure}

The cross section for the production of single top quarks and the top quark to top antiquark cross section ratio as a result of the fit are
\begin{linenomath}
\begin{align*}
\ifthenelse{\boolean{cms@external}}
{
\sigma_{t\textrm{-ch.,t}} = & 154 \pm 8\stat \pm 9\,\text{(exp)} \\ & \phantom{000} \pm 19\thy
 \pm 4\lum\unit{pb} \\
}
{
\sigma_{t\textrm{-ch.,t}} = & 154 \pm 8\stat \pm 9\,\text{(exp)} \pm 19\thy
 \pm 4\lum\unit{pb} \\
}
  = & 154 \pm 22\unit{pb}, \\
R_{t\text{-ch.}} =  & 1.81 \pm0.18\stat\pm0.15\syst .
\end{align*}
\end{linenomath}
A comparison between the measured ratio and the prediction of different PDF sets is shown in Fig.~\ref{fig:ratioplot}. With future data, this observable is expected to be sensitive to different PDF descriptions.
Using the $\sigmattop$ and $R_{t\text{-ch.}}$ measurements, the cross section of the top antiquark production is computed as
\begin{linenomath}
\begin{align*}
\ifthenelse{\boolean{cms@external}}
{
\sigma_{t\text{-ch.,}\PAQt} = & 85 \pm 10\stat \pm 4\,\text{(exp)} \\ & \phantom{00} \pm 11\thy
 \pm 2\lum\unit{pb}
 }
 {
 \sigma_{t\text{-ch.,}\PAQt} = & 85 \pm 10\stat \pm 4\,\text{(exp)} \pm 11\thy
 \pm 2\lum\unit{pb}
 }
 \\  = & 85\pm16\unit{pb},
\end{align*}
\end{linenomath}
where the uncertainties are evaluated using the correlation matrix of the simultaneous fit. This leads to the total cross section,
\begin{linenomath}
\begin{align*}
\ifthenelse{\boolean{cms@external}}
{
\sigma_{t\text{-ch.},\PQt+\PAQt} = & 238 \pm 13\stat \pm 12\,\text{(exp)} \\ & \phantom{000} \pm 26\thy
\pm 5\lum\unit{pb}
}
{
\sigma_{t\text{-ch.},\PQt+\PAQt} = & 238 \pm 13\stat \pm 12\,\text{(exp)} \pm 26\thy
\pm 5\lum\unit{pb}
}
\\ =  & 238 \pm 32 \unit{pb}.
\end{align*}
\end{linenomath}
Figure~\ref{fig:finalplot} shows the comparison of this measurement with the
standard model (SM) expectation and measurements of the single top quark $t$-channel cross section at other centre-of-mass energies.
The total cross section is used to determine the absolute value of the CKM matrix element $\abs{\mathrm{V_{tb}}}$, assuming that the other terms $\abs{\mathrm{V_{td}}}$ and $\abs{\mathrm{V_{ts}}}$ are much smaller than $\abs{\mathrm{V_{tb}}}$:
\begin{linenomath}
\begin{equation*}
\mathrm{\abs{f_{LV}V_{tb}}} = \sqrt{\frac{\sigma_{t\text{-ch.},\PQt+\PAQt}}{{\sigma^{\text{th}}_{t\text{-ch.},\PQt+\PAQt}}}},
\end{equation*}
\end{linenomath}
where $\sigma_{t\text{-ch.},\PQt+\PAQt}^{\mathrm{th}} =  \xsectheo \xsectheoscale\,\text{(scale)}\xsectheopdf \,(\mathrm{PDF} {+} \alpha_\mathrm{S})\unit{pb}$~\cite{HATOR,Kant:2014oha,Botje:2011sn} is the SM predicted value assuming $\abs{\mathrm{V_{\PQt\PQb}}} = 1$. The possible presence of an anomalous Wtb coupling is taken into account by the anomalous form factor $\mathrm{f_{LV}}$~\cite{AguilarSaavedra:2008zc}, which is 1 for the SM and deviates from 1 for physics beyond the standard model (BSM):
\begin{linenomath}
\begin{equation*}
\mathrm{\abs{f_{LV}V_{tb}}} = \vtbobs \pm \vtbobsexp\,(\text{exp}) \pm \vtbobstheo \,(\mathrm{theo}),
\end{equation*}
\end{linenomath}
where the first uncertainty contains all uncertainties on the cross section measurement, and the second uncertainty is the uncertainty on the theoretical SM prediction.
\section{Summary}
A measurement of the cross section of the $t$-channel single top quark production is presented using events with one muon and jets in the final state. The cross section for the production of single top quarks and the ratio of the top quark to top antiquark production are measured together in a simultaneous fit where the results are used to evaluate the production cross section of single top antiquarks. The measured total cross section, which currently constitutes the most precise result at 13 TeV, is used to calculate the absolute value of the CKM matrix element $\abs{\mathrm{V_{tb}}}$. All results are in agreement with recent theoretical standard model predictions.
\begin{acknowledgments}
We congratulate our colleagues in the CERN accelerator departments for the excellent performance of the LHC and thank the technical and administrative staffs at CERN and at other CMS institutes for their contributions to the success of the CMS effort. In addition, we gratefully acknowledge the computing centres and personnel of the Worldwide LHC Computing Grid for delivering so effectively the computing infrastructure essential to our analyses. Finally, we acknowledge the enduring support for the construction and operation of the LHC and the CMS detector provided by the following funding agencies: BMWFW and FWF (Austria); FNRS and FWO (Belgium); CNPq, CAPES, FAPERJ, and FAPESP (Brazil); MES (Bulgaria); CERN; CAS, MoST, and NSFC (China); COLCIENCIAS (Colombia); MSES and CSF (Croatia); RPF (Cyprus); MoER, ERC IUT and ERDF (Estonia); Academy of Finland, MEC, and HIP (Finland); CEA and CNRS/IN2P3 (France); BMBF, DFG, and HGF (Germany); GSRT (Greece); OTKA and NIH (Hungary); DAE and DST (India); IPM (Iran); SFI (Ireland); INFN (Italy); MSIP and NRF (Republic of Korea); LAS (Lithuania); MOE and UM (Malaysia); CINVESTAV, CONACYT, SEP, and UASLP-FAI (Mexico); MBIE (New Zealand); PAEC (Pakistan); MSHE and NSC (Poland); FCT (Portugal); JINR (Dubna); MON, RosAtom, RAS and RFBR (Russia); MESTD (Serbia); SEIDI and CPAN (Spain); Swiss Funding Agencies (Switzerland); MST (Taipei); ThEPCenter, IPST, STAR and NSTDA (Thailand); TUBITAK and TAEK (Turkey); NASU and SFFR (Ukraine); STFC (United Kingdom); DOE and NSF (USA).

Individuals have received support from the Marie-Curie programme and the European Research Council and EPLANET (European Union); the Leventis Foundation; the A. P. Sloan Foundation; the Alexander von Humboldt Foundation; the Belgian Federal Science Policy Office; the Fonds pour la Formation \`a la Recherche dans l'Industrie et dans l'Agriculture (FRIA-Belgium); the Agentschap voor Innovatie door Wetenschap en Technologie (IWT-Belgium); the Ministry of Education, Youth and Sports (MEYS) of the Czech Republic; the Council of Science and Industrial Research, India; the HOMING PLUS programme of the Foundation for Polish Science, cofinanced from European Union, Regional Development Fund; the Mobility Plus programme of the Ministry of Science and Higher Education (Poland); the OPUS programme of the National Science Center (Poland); the Compagnia di San Paolo (Torino); MIUR project 20108T4XTM (Italy); the Thalis and Aristeia programmes cofinanced by EU-ESF and the Greek NSRF; the National Priorities Research Program by Qatar National Research Fund; the Rachadapisek Sompot Fund for Postdoctoral Fellowship, Chulalongkorn University (Thailand); the Chulalongkorn Academic into Its 2nd Century Project Advancement Project (Thailand); and the Welch Foundation, contract C-1845.

\end{acknowledgments}
\bibliography{auto_generated}
\cleardoublepage \appendix\section{The CMS Collaboration \label{app:collab}}\begin{sloppypar}\hyphenpenalty=5000\widowpenalty=500\clubpenalty=5000\input{TOP-16-003-authorlist.tex}\end{sloppypar}
\end{document}

%% file: TOP-16-003-authorlist.tex
\textbf{Yerevan Physics Institute,  Yerevan,  Armenia}\\*[0pt]
A.M.~Sirunyan, A.~Tumasyan
\vskip\cmsinstskip
\textbf{Institut f\"{u}r Hochenergiephysik,  Wien,  Austria}\\*[0pt]
W.~Adam, E.~Asilar, T.~Bergauer, J.~Brandstetter, E.~Brondolin, M.~Dragicevic, J.~Er\"{o}, M.~Flechl, M.~Friedl, R.~Fr\"{u}hwirth\cmsAuthorMark{1}, V.M.~Ghete, C.~Hartl, N.~H\"{o}rmann, J.~Hrubec, M.~Jeitler\cmsAuthorMark{1}, A.~K\"{o}nig, I.~Kr\"{a}tschmer, D.~Liko, T.~Matsushita, I.~Mikulec, D.~Rabady, N.~Rad, B.~Rahbaran, H.~Rohringer, J.~Schieck\cmsAuthorMark{1}, J.~Strauss, W.~Waltenberger, C.-E.~Wulz\cmsAuthorMark{1}
% ...
\vskip\cmsinstskip
% ...
1:~~Also at Vienna University of Technology, Vienna, Austria\\
% ...
